\documentclass[JHEP,12pt]{article}
\usepackage{amssymb,amsmath}
\usepackage{braket}
\usepackage{jheppub}
\usepackage{bbold}
\usepackage{comment}

\author{Raphael Bousso, Venkatesa Chandrasekaran, Pratik Rath, \\ and Arvin Shahbazi-Moghaddam}

\affiliation{Center for Theoretical Physics and Department of Physics,\\
University of California, Berkeley, CA 94720, U.S.A. and \\
Lawrence Berkeley National Laboratory, Berkeley, CA 94720, U.S.A.} 

\emailAdd{bousso@berkeley.edu}
\emailAdd{ven\_chandrasekaran@berkeley.edu}
\emailAdd{pratik\_rath@berkeley.edu}
\emailAdd{arvinshm@berkeley.edu}

\title{Gravity Dual of Connes Cocycle Flow}
\abstract{We define the ``kink transform'' as a one-sided boost of bulk initial data about the Ryu-Takayanagi surface of a boundary cut. For a flat cut, we conjecture that the resulting Wheeler-DeWitt patch is the bulk dual to the boundary state obtained by Connes cocycle (CC) flow across the cut. The bulk patch is glued to a precursor slice related to the original boundary slice by a one-sided boost. This evades ultraviolet divergences and distinguishes our construction from one-sided modular flow. We verify that the kink transform is consistent with known properties of operator expectation values and subregion entropies under CC flow. CC flow generates a stress tensor shock at the cut, controlled by a shape derivative of the entropy; the kink transform reproduces this shock holographically by creating a bulk Weyl tensor shock. We also go beyond known properties of CC flow by deriving novel shock components from the kink transform.}

\begin{document}
\maketitle

\section{Introduction}
\label{intro}

The AdS/CFT duality~\cite{Maldacena:1997re,Witten:1998qj,Gubser:1998bc} has led to tremendous progress in the study of quantum gravity. However, our understanding of the holographic dictionary remains limited. In recent years, quantum error correction was found to play an important role in the emergence of a gravitating (``bulk'') spacetime from the boundary theory~\cite{Almheiri:2014lwa,Harlow:2016vwg,Hayden:2018khn}. The study of modular operators led to the result that the boundary relative entropy in a region $A$ equals the bulk relative entropy in its entanglement wedge $\text{EW}(A)$~\cite{Jafferis:2015del}.
The combination of these insights was used to derive subregion duality: bulk operators in $\text{EW}(A)$ can in principle be reconstructed from operators in the subregion $A$~\cite{Dong:2016eik}.

The relation between bulk modular flow in $\text{EW}(A)$ and boundary modular flow in $A$ has been used to explicitly reconstruct bulk operators both directly~\cite{Faulkner:2017vdd,Chen:2019iro}, and indirectly via the Petz recovery map and its variants~\cite{Cotler:2017erl,Chen:2019gbt,Penington:2019kki}. Thus, modular flow has shed light on the emergence of the bulk spacetime from entanglement properties of the boundary theory.

Modular flow has also played an important role in proving various properties of quantum field theory (QFT), such as the averaged null energy condition (ANEC) and quantum null energy condition (QNEC) \cite{Faulkner:2016mzt,Balakrishnan:2017bjg}. Tomita-Takesaki theory, the study of modular flow in algebraic QFT, puts constraints on correlation functions that are otherwise hard to prove directly \cite{Lashkari:2018nsl}. 

Recently, an alternate proof of the QNEC was found using techniques from Tomita-Takesaki theory \cite{Ceyhan:2018zfg}. The key ingredient was Connes cocyle (CC) flow. Given a subregion $A$ and global pure state $\psi$, Connes cocycle flow acts with a certain combination of modular operators to generate a sequence of well-defined global states $\psi_s$. In the limit $s \rightarrow \infty$, these states saturate Wall's ``ant conjecture''~\cite{Wall:2017blw} on the minimum amount of energy in the complementary region $A'$. This proves the ant conjecture, which, in turn, implies the QNEC.

CC flow also arises from a fascinating interplay between quantum gravity, quantum information, and QFT. Recently, the classical black hole coarse-graining construction of Engelhardt and Wall~\cite{Engelhardt:2017aux} was conjecturally extended to the semiclassical level~\cite{Bousso:2019dxk}. In the non-gravitational limit, this conjecture requires the existence of flat space QFT states with properties identical to the $s\to\infty$ limit of CC flowed states. This is somewhat reminiscent of how the QNEC was first discovered as the nongravitational limit of the quantum focusing conjecture \cite{Bousso:2015wca}. Clearly, CC flow plays an important role in the connection between QFT and gravity. Our goal in this paper is to investigate this connection at a deeper level within the setting of AdS/CFT.

In Sec.~\ref{CC}, we define CC flow and discuss some of its properties. If $\partial A$ lies on a null plane in Minkowski space, operator expectation values and subregion entropies within the region $A$ remain the same, whereas those in $A'$ transform analogously to a boost~\cite{Ceyhan:2018zfg}. Further, CC flowed states $\psi_s$ exhibit a characteristic stress tensor shock at the cut $\partial A$, controlled by the derivative of the von Neumann entropy of the region $A$ in the state $\psi$ under shape deformations of $\partial A$~\cite{Bousso:2019dxk}.

As is familiar from other examples in holography, bulk duals of complicated boundary objects are often much simpler \cite{Ryu:2006bv,Koeller:2015qmn}.
Motivated by the known properties of CC flow, we define a bulk construction in Sec.~\ref{kink}, which we call the ``kink transform.'' This is a one-parameter transformation of the initial data of the bulk spacetime dual to the original boundary state $\psi$. We consider a Cauchy surface $\Sigma$ that contains the Ryu-Takayanagi surface $\mathcal{R}$ of the subregion $A$. The kink transform acts as the identity except at $\mathcal{R}$, where an $s$-dependent shock is added to the extrinsic curvature of $\Sigma$. We show that this is equivalent to a one-sided boost of $\Sigma$ in the normal bundle to $\mathcal{R}$. We prove that the new initial data satisfies the gravitational constraint equations, thus demonstrating that the kink transform defines a valid bulk spacetime $\mathcal{M}_s$. We show that $\mathcal{M}_s$ is independent of the choice of $\Sigma$. 

We propose that $\mathcal{M}_s$ is the holographic dual to the CC-flowed state $\psi_s$, if the boundary cut $\partial A$ is (conformally) a flat plane in Minkowski space.

In Sec.~\ref{duality}, we provide evidence for this proposal. The kink transform separately preserves the entanglement wedges of $A$ and $A'$, but it glues them together with a relative boost by rapidity $2\pi s$. This implies the one-sided expectation values and subregion entropies of the CC flowed state $\psi_s$ are correctly reproduced when they are computed holographically in the bulk spacetime $\mathcal{M}_s$. We then perform a more nontrivial check of this proposal. By computing the boundary stress tensor holographically in $\mathcal{M}_s$, we reproduce the stress tensor shock at $\partial A$ in the CC-flowed state $\psi_s$. 

Having provided evidence for kink transform/CC flow duality, we use the duality to make a novel prediction for CC flow in Sec.~\ref{prediction}. The kink transform fully determines all independent components of the shock at $\partial A$ in terms of shape derivatives of the entanglement entropy. Strictly, our results only apply only to the CC flow of a holographic CFT across a planar cut. However, their universal form suggests that they will hold for general QFTs under CC flow. Moreover, the shocks we find agree with properties required to exist in quantum states under the coarse-graining proposal of Ref.~\cite{deBoer:2019uem}. Thus, our new results may also hold for CC flow across general cuts of a null plane.

In Sec.~\ref{discussion}, we discuss the relation of our construction to earlier work on the role of modular flow in AdS/CFT~\cite{Jafferis:2014lza,Jafferis:2015del,Faulkner:2018faa}.
The result of Jafferis {\em et al.\ }(JLMS)~\cite{Jafferis:2015del} has conventionally been understood as a relation that holds for a small code subspace of bulk states on a fixed background spacetime.
However, results from quantum error correction suggest that this code subspace could be made much larger to include different background geometries \cite{Harlow:2016vwg,Akers:2018fow,Dong:2018seb,Dong:2019piw}.
Our proposal then follows from such an extended version of the JLMS result which includes non-perturbatively different background geometries. Equipped with this understanding, we can distinguish our proposal from the closely related bulk duals of one-sided modular flowed states~\cite{Jafferis:2014lza,Faulkner:2018faa}.
We provide additional evidence for our proposal based on two sided correlation functions of heavy operators, and we discuss generalizations and applications of the proposed kink transform/CC flow duality.

In Appendix~\ref{nulllimit} we derive the null limit of the kink transform, and show that it generates a Weyl shock, which provides intuition for how the kink transform modifies gravitational observables.

\section{Connes Cocycle Flow}
\label{CC}

In this section, we review Connes cocycle flow and its salient properties; for more details see \cite{Ceyhan:2018zfg, Bousso:2019dxk}. We then reformulate Connes cocycle flow in as a simpler map to a state defined on a ``precursor" slice. This will prove useful in later sections.    

\subsection{Definition and General Properties}
\label{general}

Consider a quantum field theory on Minkowski space $\mathbb{R}^{d-1,1}$ in standard Cartesian coordinates $(t, x,y_1,\ldots,y_{d-2})$. Consider a Cauchy surface $\mathcal{C}$ that is the disjoint union of the open regions $A_0, A'_0$ and their shared boundary $\partial A_0$. Let $\mathcal{A}_0, \mathcal{A}'_0$ denote the associated algebras of operators. Let $|\psi\rangle$ be a cyclic and separating state on $\mathcal{C}$, and denote by $|\Omega\rangle$ the global vacuum (the assumption of cyclic and separating could be relaxed for $|\psi\rangle$, at the cost of complicating the discussion below). The Tomita operator is defined by 
\begin{align}
    S_{\psi|\Omega; \mathcal{A}_0}\alpha |\psi\rangle = \alpha^{\dagger}|\Omega\rangle, \forall \alpha \in \mathcal{A}_0 ~.
\end{align}
The relative modular operator is defined as
\begin{align}
    \Delta_{\psi|\Omega%; \mathcal{A}_0
    } \equiv S^{\dagger}_{\psi|\Omega;\mathcal{A}_0}S_{\psi|\Omega;\mathcal{A}_0}~,
\end{align}
and the vacuum modular operator is
\begin{equation}
    \Delta_\Omega \equiv \Delta_{\Omega|\Omega}~.
\end{equation}
Note that we do not include the subscript $\mathcal{A}_0$ on $\Delta$; instead, for modular operators, we indicate whether they were constructed from $\mathcal{A}_0$ or $\mathcal{A}'_0$ by writing $\Delta$ or $\Delta'$. 

Connes cocycle (CC) flow of $|\psi\rangle$ generates a one parameter family of states $|\psi_s\rangle$, $s\in \mathbb{R}$,  defined by 
\begin{equation}
    |\psi_s\rangle = (\Delta'_{\Omega})^{is}(\Delta'_{\Omega|\psi})^{-is}|\psi\rangle ~.
    \label{ccflow}
\end{equation}
Thus far the definitions have been purely algebraic. In order to elucidate the intuition behind CC flow, let us write out the modular operators in terms of the left and right density operators, $\rho^{\psi}_{A_0} = \text{Tr}_{A'_0}|\psi\rangle\langle \psi|$ and $\rho^{\psi}_{A'_0} =\text{Tr}_{A_0}|\psi\rangle\langle \psi|$:\footnote{We follow the conventions in \cite{Witten:2018zxz} where complement operators are written to the right of the tensor product.} 
\begin{align}
    \Delta_{\psi|\Omega%; A_0
    } &= \rho^{\Omega}_{A_0}\otimes (\rho^{\psi}_{A'_0})^{-1} ~.
\end{align} 
One finds that the CC operator acts only in $\mathcal{A}'_0$:
\begin{equation}\label{eq:CCrho}
    (\Delta'_{\Omega})^{is}(\Delta'_{\Omega|\psi})^{-is} = (\rho^\Omega_{A'_0})^{is} (\rho^{\psi}_{A'_0})^{-is}\in \mathcal{A}'_0~.
\end{equation}
It follows that the reduced state on the right algebra satisfies
\begin{equation}
    \rho^{\psi_s}_{A_0} = \rho^{\psi}_{A_0}~.\label{sameright}
\end{equation}
Therefore, expectation values of observables $\mathcal{O}\in \mathcal{A}_0$ remain invariant under CC flow. These heuristic arguments would be valid only for finite-dimensional Hilbert spaces~\cite{Witten:2018zxz}; but Eq.~\eqref{eq:CCrho} can be derived rigorously~\cite{Ceyhan:2018zfg}.

It can also be shown that $(\Delta'_{\psi|\Omega})^{is}\Delta^{is}_{\Omega|\psi} = 1$. Hence for operators $\mathcal{O}'\in \mathcal{A}'_0$, one finds that CC flow acts as $\Delta_{\Omega}^{is}$ inside of expectation values:
\begin{align}
    \langle \psi_s| \mathcal{O}'|\psi_s\rangle &= \text{Tr}_{\mathcal{A}'_0}\left[\rho^{\psi}_{A'_0}
    (\Delta_{\psi|\Omega}^{-is}\Delta_{\Omega}^{is})
    \mathcal{O}' (\Delta_{\Omega}^{-is}\Delta^{is}_{\psi|\Omega})\right] ~, \nonumber 
    \\ &= \text{Tr}_{\mathcal{A}'_0}\left(\rho^{\psi}_{A'_0}(\rho^{\Omega}_{A'_0})^{-is}
    \mathcal{O}'(\rho^{\Omega}_{A'_0})^{is}\right) 
    \\ &= \text{Tr}\left[|\psi\rangle\langle \psi|\Delta_{\Omega}^{is} ({\mathbf 1}\otimes \mathcal{O}')\Delta_{\Omega}^{-is}\right] ~,
\end{align}
where we have used the cyclicity of the trace. 

To summarize, expectation values of one-sided operators transform as follows:
\begin{align}
\langle \psi_s| \mathcal{O}|\psi_s\rangle &= \langle \psi| \mathcal{O}|\psi\rangle~,  \\ 
    \langle \psi_s| \mathcal{O}'|\psi_s\rangle &= \langle \psi|\Delta^{is}_{\Omega}\mathcal{O'}\Delta^{-is}_{\Omega}|\psi\rangle ~. \label{ccobs}
\end{align}
There is no simple description of two-sided correlators in $|\psi_s\rangle$ such as $\langle \psi_s|\mathcal{O}\mathcal{O'}|\psi_s\rangle$; we discuss such objects in Sec.~\ref{other}.

\subsection{CC Flow from Cuts on a Null Plane}
\label{nullcuts}

Let us now specialize to the case where $\partial A_0$ corresponds to a cut $v = V_0(y)$ of the Rindler horizon $u = 0$. We have introduced null coordinates $u = t-x$ and $v = t+x$ and denoted the transverse coordinates collectively by $y$. It can be shown that the modular operator $\Delta^{is}_\Omega$ acts locally on each null generator $y$ of $u=0$ as a boost about the cut $V_0(y)$~\cite{Casini:2017roe}. More explicitly, one can define the \emph{full} vacuum modular Hamiltonian $\widehat{K}_{V_0}$ by 
\begin{align}
    \widehat{K}_{V_0} = -\log \Delta_{\Omega; \mathcal{A}_{V_0}}~.
\end{align}
We can write the full modular Hamiltonian as 
\begin{align}
    \widehat{K}_{V_0} =  K_{V_0}\otimes\boldsymbol{1}' - \boldsymbol{1}\otimes K'_{V_0}~.
\end{align}
Let $\Delta$ denote vacuum subtraction, $\Delta \langle \mathcal{O}\rangle = \langle \mathcal{O}\rangle_{\psi}-\langle \mathcal{O}\rangle_{\Omega}$. Then, for arbitrary cuts of the Rindler horizon, we have \cite{Casini:2017roe}
\begin{align}
    \Delta \langle K'_{V_0}\rangle = -2\pi \int dy \int_{-\infty}^{V_0}dv[v-V_0(y)]\langle T_{vv}\rangle_{\psi} ~,\label{modham1}
\end{align} 
and similarly for $K_{V_0}$. Thus $K'_{V_0}$ is simply the boost generator about the cut $V_0(y)$ in the left Rindler wedge. That is, it generates a $y$-dependent dilation,
\begin{equation}
    v\to V_0(y) + [v-V_0(y)]e^{ 2\pi s} ~.
\end{equation}
%Of particular interest are the subset of observables localized to the Rindler horizon $u = 0$ in standard lightcone coordinates $\left(u,v,y^1,\ldots,y^{d-2}\right)$. Let $V_0$ denote the flat cut in lightcone coordinates. 
This allows us to evaluate Eq.~\eqref{ccobs} explicitly for local operators at $u=0$. For example, the CC flow of the stress tensor is  
\begin{align}
    \langle \psi_s| T_{vv}|\psi_s\rangle\lvert_{v < V_0} = e^{-4\pi s}\langle \psi| T_{vv}\left(V_0 + e^{-2\pi s}(v-V_0) \right)|\psi\rangle\lvert_{v < V_0}~, \label{stresstransform}
\end{align}
and similarly for the other components of $T_{\mu \nu}$. There is a slight caveat here since $\Delta^{is}_{\Omega}$ only acts as a boost strictly at $u = 0$. This would be sufficient for free theories, where $T_{vv}$ can be defined through null quantization on the Rindler horizon with a smearing that only needs support on $u = 0$ \cite{Wall:2011hj}. More generally, $T_{\mu\nu}$ must be smeared in an open neighborhood of $u=0$. However, if $V_0(y)$ is a perturbation of a flat cut then one can show that inside correlation functions $\Delta^{is}_\Omega$ approximately acts as a boost with subleading errors that vanish as $u\rightarrow 0$, to all orders in the perturbation \cite{Balakrishnan:2017bjg, Balakrishnan:2020lbp}. In the non-perturbative case, evidence comes from the fact that classically the vector field on the Rindler horizon which generates boosts about $V_0(y)$ can be extended to an approximate Killing vector field in a neighborhood of the horizon \cite{Koga:2001vq, Ashtekar:2001jb}. Therefore we expect Eq.~\eqref{stresstransform} to hold on the null surface even after smearing. 

Now consider a second cut $V(y)$ of the Rindler horizon which lies entirely below $V_0(y)$, so $V<V_0$ for all $y$. The cut defines a surface $\partial A_V$ that splits a Cauchy surface $\mathcal{C}_V = A'_V \cup \partial A_V \cup A_V$; we take $A'_V$ to be the ``left" side ($v<V$), with operator algebra $\mathcal{A}'_V$. The Araki definition of relative entropy is \cite{Witten:2018zxz} 
\begin{align}
S'_{\text{rel}}(\psi|\Omega; V) = -\langle \psi| \log\Delta_{\psi| \Omega; \mathcal{A}'_V}|\psi\rangle~.
\end{align}
It has the following transformation properties \cite{Ceyhan:2018zfg}:
\begin{align}
    S_{\text{rel}}(\psi_s|\Omega; V) &= S_{\text{rel}}(\psi|\Omega; V_0 + e^{-2\pi s}(V-V_0) ) ~,\label{reltrans}  \\ 
    \frac{\delta S_{\text{rel}}(\psi_s|\Omega; V)}{\delta V} &= e^{-2\pi s}\frac{\delta S_{\text{rel}}(\psi|\Omega; V_0 + e^{-2\pi s}(V-V_0))}{\delta V}~. 
    \label{reldertrans}
\end{align}
Moreover, the ``left" von Neumann entropy is defined as
\begin{align}
S'(\psi, V) = -\text{tr}_{A'_V}{\rho^{\psi}_{A'_V} \log \rho^{\psi}_{A'_V}}~.
\end{align}
With these definitions in hand, one can decompose the relative entropy as 
\begin{align}
    S'_{\text{rel}}(\psi|\Omega; V) = \Delta \langle K'_V\rangle - \Delta S'(V) ~. \label{relentropy}
\end{align}
At this point we drop the explicit vacuum subtractions, as we will only be interested in shape derivatives of the vacuum subtracted quantities, which automatically annihilate the vacuum expectation values. In particular, one can directly compute shape derivatives of $K'_V$:
\begin{align}
    \frac{\delta \langle K'_V\rangle_{\psi}}{\delta V}\Big \lvert_{V_0} = 2\pi \int_{-\infty}^{V_0}dv \ \langle T_{vv}\rangle_{\psi} ~. \label{modham}
\end{align}
Hence the transformations of both $K'_V$ and its derivative simply follow from Eq.~\eqref{stresstransform}. 

Combining Eq.~\eqref{reltrans} and Eq.~\eqref{modham1}, as well as Eq.~\eqref{reldertrans} and Eq.~\eqref{modham}, we see that $S'(\psi,V)$ and its derivative transform as
\begin{align}
S'(\psi_s,V) &= S'(\psi,V_0 + e^{-2\pi s}(V-V_0)) ~, \\  
\frac{\delta S'}{\delta V}\Big\lvert_{\psi_s,V} &= e^{-2\pi s}\frac{\delta S'}{\delta V}\Big\lvert_{\psi,V_0 + e^{-2\pi s}(V-V_0)}~. \label{entropytransform}
\end{align}
The respective properties of the complement entropy follow from purity. 

\subsection{Stress Tensor Shock at the Cut}
\label{stsc}

CC flow generates a stress tensor shock at the cut $V_0$, proportional to the jump in the variation of the one-sided von Neumann entropy under deformations, at the cut~\cite{Bousso:2019dxk}. To see this, let us start with the sum rule derived in~\cite{Ceyhan:2018zfg} for null variations of relative entropy:\footnote{For type I algebras, one can derive the analogous sum rule from simpler arguments~\cite{Bousso:2014uxa}.}
\begin{align}
    2\pi(P_s - e^{-2\pi s}P_0) = (e^{-2\pi s}-1)\frac{\delta S'_{\text{rel}}(\psi|\Omega; V)}{\delta V}\Big\lvert_{V_0}~,
    \label{sumrule}
\end{align}
where 
\begin{align}
    P \equiv \int_{-\infty}^{\infty}dv\   T_{vv}
\end{align}
is the averaged null energy operator at $u=0$, and $P_s \equiv \langle \psi_s| P|\psi_s\rangle$, so in particular $P_0 \equiv \langle \psi| P|\psi\rangle$. (There is one such operator for every generator, {\em i.e.}, for every $y$.) 

Inserting Eq.~\eqref{relentropy} and Eq.~\eqref{modham} into Eq.~\eqref{sumrule}, and making use of Eq.~\eqref{stresstransform}, we see that there must exist a shock at $v=V_0(y)$:
\begin{align}
    \langle \psi_s| T_{vv}|\psi_s\rangle = (1-e^{-2\pi s})\frac{1}{2\pi}\frac{\delta S'}{\delta V}\Big \lvert_{V_0}\delta(v-V_0) + o(\delta)~.\label{boundary shock}
\end{align}
Here $o(\delta)$ designates the finite (non-distributional) terms. These are determined by Eq.~\eqref{stresstransform}, and by its trivial counterpart in the $v>V_0$ region.

This $s$-dependent shock is a detailed characteristic of the CC flowed state. As such, reproducing it through the holographic dictionary will be the key test of our proposal of the bulk dual of CC flow (see Sec.~\ref{duality}).

\subsection{Flat Cuts and the Precursor Slice}\label{precursor}

For the remainder of the paper we further specialize to flat cuts of the Rindler horizon, so that $\partial A_0$ corresponds to $u = v = 0$. We therefore set $V_0 = 0$ in what follows. We take $\mathcal{C}$ to be the Cauchy surface $t=0$, so that $A_0$ ($t=0$, $x>0$) and $A_0'$ ($t=0$, $x<0$) are partial Cauchy surfaces for the right and left Rindler wedges. 

In this case $\Delta^{is}_{\Omega}$ is a global boost by rapidity $s$ about $\partial A_0$ \cite{Bisognano:1976za}. Thus, it has a simple geometric action not only on the null plane $u=0$, but everywhere.
CC flow transforms observables in $\mathcal{A}'_0$ by $\Delta^{is}_{\Omega}$ and leaves invariant those in $\mathcal{A}_0$. For a flat cut, this action can be represented as a geometric boost in the entire left Rindler wedge. This allows us to characterize the CC flowed state $|\psi(s)\rangle$ on $\mathcal{C}$ very simply in terms of a  different state on a different Cauchy surface which we call the ``precursor slice''. This description will motivate the formulation of our bulk construction in Sec. \ref{formulation}.

By Eq.~\eqref{ccobs}, the CC flowed state on the slice $\mathcal{C}$,
\begin{align}
\ket{\psi_s(\mathcal{C})} = (\Delta'_{\Omega})^{is}(\Delta'_{\Omega|\psi})^{-is} \ket{\psi(\mathcal{C})}~,
\end{align}
satisfies
\begin{align}
&\bra{\psi_s(\mathcal{C})} \mathcal{O}_{A}  \ket{\psi_s(\mathcal{C})} = \bra{\psi(\mathcal{C})} \mathcal{O}_{A}  \ket{\psi(\mathcal{C})}~, \label{CCR-passive}  \\
&\bra{\psi_s(\mathcal{C})} \Delta_{\Omega}^{-is} \mathcal{O}_{A'} \Delta_{\Omega}^{is} \ket{\psi_s(\mathcal{C})} = \bra{\psi(\mathcal{C})} \mathcal{O}_{A'} \ket{\psi(\mathcal{C})} ~,\label{CCL-passive}
\end{align}
where $\mathcal{O}_{A}$ and $\mathcal{O}_{A'}$ denote an arbitrary collection of local operators that act on spacelike half-slices $A$ and $A'$ of $\mathcal{C}$ respectively.\footnote{More precisely, one would have to smear the operator in a codimension 0 neighborhood of points on the slices.} In the second equality above, we used the fact that $\Delta_{\Omega}^{is}$ acts as a global boost to move it to the other side of the equality, compared to Eq.~\eqref{ccobs}.

We work in the Schr\"odinger picture where the argument $\mathcal{C}$ should be interpreted as the time variable. The fact that $\Delta_{\Omega}^{is}$ acts as a boost around $\partial A_{0}$ motivates us to consider the time slice
\begin{equation}
    \mathcal{C}_s = A'_s\cup \partial A_0\cup A_0~,
\end{equation}
where 
\begin{equation}
    A'_{s} =  \{t=(\tanh 2\pi s) x , ~x<0\} ~.
    \label{aps}
\end{equation}
By Eqs.~\eqref{CCR-passive} and \eqref{CCL-passive}, each side of the CC-flowed state $\ket{\psi_s(\mathcal{C}_{s})}$ is simply related to the left and right restrictions of the original state on the original slice:
\begin{align}
&\bra{\psi_s(\mathcal{C}_{s})} \mathcal{O}_{A}  \ket{\psi_s(\mathcal{C}_{s})} = \bra{\psi(\mathcal{C})} \mathcal{O}_{A}  \ket{\psi(\mathcal{C})}~, \label{precursorR} \\
&\bra{\psi_s(\mathcal{C}_{s})} \mathcal{O}_{A'_{s}}  \ket{\psi_s(\mathcal{C}_{s})} = \bra{\psi(\mathcal{C})} \mathcal{O}_{A'} \ket{\psi(\mathcal{C})}~. \label{precursorL}
\end{align}
In the second equation, $\mathcal{O}_{A'_{s}}$ denotes local operators on $A'_{s}$ which are analogous to $\mathcal{O}_{A'}$ on $A'$. More precisely, because the intrinsic metric of $A'$ and $A'_{s}$ are the same, there exists a natural map between local operators on $A'$ and $A'_{s}$. 

In words, Eqs.~\eqref{precursorR} and \eqref{precursorL} say that correlation functions in each half of $\mathcal{C}$ in the state $\ket{\psi(\mathcal{C})}$ are equal to the analogous correlation functions on each half of $\mathcal{C}_{s}$ in the state $\ket{\psi_s(\mathcal{C}_s)}$. This justifies calling $\mathcal{C}_s$ the precursor slice since the CC flowed state on $\mathcal{C}$ arises from it by time evolution.

We find it instructive to repeat this point in the less rigorous language of density operators. In the density operator form of CC flow,
\begin{align}
\ket{\psi_s(\mathcal{C})} = (\rho^{\Omega}_{A'_0})^{i s}(\rho^{\psi}_{A'_0})^{-i s} \ket{\psi(\mathcal{C})}~,
\label{ccdo}
\end{align}
it is evident that the action of $(\rho^{\Omega}_{A'_0})^{i s}$ can be absorbed into a change of time slice $\mathcal{C} \to \mathcal{C}_{s}$:
\begin{align}\label{precursor-rho}
\ket{\psi_{s}(\mathcal{C}_{s})} = (\rho^{\psi}_{A'_0})^{-i s} \ket{\psi(\mathcal{C})}~.
\end{align}
Tracing out each side of $\partial A_{0}$ implies
\begin{align}
&\rho^{\psi_{s}}_{A_0}= \rho^{\psi}_{A_0}~,
\label{CCkink11}\\
&\rho^{\psi_{s}}_{A'_{s}} =  \rho^{\psi}_{A'_0}\label{CCkink22}~.
\end{align}
The first equality is trivial and was already discussed in Eq.~\eqref{sameright}. The second equality follows because $(\rho^{\psi}_{A'_0})^{i s}$ commutes with $(\rho^{\psi}_{A'_0})$. This is the density operator version of Eqs.~\eqref{precursorR} and \eqref{precursorL}.

Eq.~\eqref{precursor-rho} should be contrasted with the one-sided modular-flowed state $\ket{\phi(\mathcal{C})}= (\rho^{\psi}_{A'_0})^{-i s} \ket{\psi(\mathcal{C})}$. The latter state would live on the original slice $\mathcal{C}$, but it is not well-defined since it would have infinite energy at the entangling surface.

It will be useful to define new coordinates adapted to the precursor slice $\mathcal{C}_s$. Let
\begin{align}
    &\tilde{v} = v\,\Theta(v) +e^{-2\pi s}\,v\,(1-\Theta(v))~,\\
    &\tilde{u} = e^{2\pi s}u\,\Theta(u) +\,u\,(1-\Theta(u))~,
    \label{tildecoords}
\end{align}
where $\Theta(.)$ is the Heaviside step function. Let $\tilde{t} = \frac{1}{2}(\tilde{v}+\tilde{u})$ and $\tilde{x} = \frac{1}{2}(\tilde{v}-\tilde{u})$. In these coordinates, the Minkowski metric takes the form
\begin{align}
    ds^2 &= \left[\Theta(\tilde{t}+\tilde{x})+e^{2\pi s}(1-\Theta(\tilde{t}+\tilde{x})) \right] \left[e^{-2\pi s}\Theta(\tilde{t}-\tilde{x})+(1-\Theta(\tilde{t}-\tilde{x})) \right] (-d\tilde{t}^2 + d\tilde{x}^2)\nonumber\\
    &+ d^{d-2}y ~, \label{tildeeta}
\end{align}
and the precursor slice corresponds to $\tilde t=0$.

In these ``tilde" coordinates, the stress tensor shock of Eq.~\eqref{boundary shock} takes the form\footnote{We remind the reader that $o(\delta)$ refers to any finite (non-distributional) terms.}
\begin{align}
    \langle \psi_s| T_{\tilde{v}\tilde{v}}|\psi_s \rangle = \frac{1}{2\pi}\left(\frac{\partial v}{\partial \tilde{v}} \right)^2(1-e^{-2\pi s})\frac{\delta S}{\delta V}\Big\lvert_{V=0}\delta(v) +o(\delta) \label{transform}~.
\end{align}
Recall that the entropy variation is evaluated in the state $|\psi\rangle$. By Eq.~\eqref{entropytransform}, 
\begin{align}
    \frac{\delta S}{\delta V}\Big \lvert_{\psi} = \frac{\delta S}{\delta \widetilde{V}}\Big \lvert_{\psi_s}~,
\end{align}
where $\widetilde{V}(y)$ is a cut of the Rindler horizon in the $\tilde{v}$ coordinates. Thus we may instead evaluate the entropy variation in the state $|\psi_s\rangle$ on the precursor slice. This will be convenient when matching the bulk and boundary. 

The Jacobian in Eq.~\eqref{transform} has a step function in it, as will the Jacobian coming from $\delta(v)$. A step function multiplying a delta function is well-defined if one averages the left and right derivatives: 
\begin{align}\label{nullshock}
    \left(\frac{\partial v}{\partial \tilde{v}} \right)^2\delta(v) = \frac{1}{2}\left(\frac{\partial v}{\partial \tilde{v}}\Big \lvert_{0^-} +\frac{\partial v}{\partial \tilde{v}}\Big \lvert_{0^+}  \right)\delta(\tilde{v}) ~.
\end{align}
Thus Eq.~\eqref{transform} becomes
\begin{align}
    \langle \psi_s| T_{\tilde{v}\tilde{v}}(\tilde v)|\psi_s\rangle = \frac{1}{2\pi}\sinh(2\pi s)\frac{\delta S}{\delta \widetilde{V}}\Big \lvert_{\psi_s,\widetilde{V}=0}\delta(\tilde{v})+ o(\delta). \label{shocktransform}
\end{align}

Since we are dealing with a flat cut, the symmetry $s \leftrightarrow -s, v \leftrightarrow u$ implies that CC flow also generates a $T_{uu}$ shock in the state $|\psi_s\rangle$ at $u = v = 0$:
\begin{align}\label{Ttutu}
    \langle \psi_s| T_{\tilde{u}\tilde{u}}|\psi_s\rangle = \frac{1}{2\pi}\sinh(2\pi s)\frac{\delta S}{\delta \widetilde{U}}\Big \lvert_{\psi_s,\widetilde{V}=0}\delta(\tilde{u})+o(\delta)~.
\end{align}
(Note that $\delta/\delta V$ goes to $-\delta/\delta U$.) The linear combination
\begin{align}
    \langle \psi_s|T_{\tilde{t}\tilde{x}}(\tilde t, \tilde x)|\psi_s\rangle & = \frac{1}{2\pi}\sinh(2\pi s)\frac{\delta S}{\delta \widetilde{X}}\Big \lvert_{\psi_s,\widetilde{X}=0}\delta(\tilde{x})+ \langle \psi| T_{tx}(t=\tilde t,x=\tilde x)|\psi\rangle~. \label{spacelikeshock}
\end{align}
will be useful in Sec.~\ref{duality}. The last term was obtained from Eqs.~\eqref{CCkink11} and \eqref{CCkink22}; it makes the finite piece explicit. Note that these equations are valid in the entire left and right wedges, not just on $\mathcal{C}_s$.

\section{Kink Transform}
\label{kink}

In this section, we introduce a novel geometric transformation called the kink transform. The construction is motivated by thinking about what the bulk dual of the boundary CC flow would be in the context of AdS/CFT.  As we discussed in Sec. \ref{CC}, CC flow boosts observables in $D(A')$ and leaves observables in $D(A)$ unchanged. Subregion duality in AdS/CFT then implies that the bulk dual of the state $\ket{\psi_s}$ has to have the property that the entanglement wedges of $D(A)$ and $D(A')$ will be diffeomorphic to those of the state $\ket{\psi}$, but are glued together with a ``one-sided boost" at the HRT surface. In a general geometry, a boost Killing symmetry need not exist. The kink transform appropriately generalizes the notion of a one-sided boost to any extremal surface.

In Sec.~\ref{formulation}, we formulate the kink transform. In Sec.~\ref{properties}, we describe a different but equivalent formulation of the kink transform and show that the kink transform results in the same new spacetime, regardless of which Cauchy surface containing the extremal surface is used for the construction. In Sec.~\ref{duality}, we will describe the duality between the bulk kink transform and the boundary CC flow in AdS/CFT and provide evidence for it.

\subsection{Formulation}\label{formulation}

Consider a $d+1$ dimensional spacetime $\mathcal{M}$ with metric $g_{\mu\nu}$ satisfying the Einstein field equations. (We will discuss higher curvature gravity in Sec.~\ref{hcc}.) Let $\Sigma$ be a Cauchy surface of $\mathcal{M}$ that contains an extremal surface $\mathcal{R}$ of codimension 1 in $\Sigma$. (That is, the expansion of both sets of null geodesics orthogonal to $\mathcal{R}$ vanishes.)

Initial data on $\Sigma$ consist of~\cite{wald2010general} the intrinsic metric $(h_{\Sigma})_{ab} $ and the extrinsic curvature,
\begin{align}\label{extrinsic}
(K_{\Sigma})_{ab} =P^{\mu}_{a} P^{\nu}_{b} \nabla_{(\mu} t_{\nu)}~.
\end{align}
Here $P^{\mu}_{a}$ is the projector from $\mathcal{M}$ onto $\Sigma$, and $t^{\mu}$ is the unit norm timelike vector field orthogonal to $\Sigma$. Indices $a,b,\ldots$ are reserved for directions tangent to $\Sigma$. For matter fields, initial data consist of the fields and normal derivatives, for example $\phi(w^a)$ and $[t^{\mu} \nabla_{\mu}\phi](w^a)$, where $\phi$ is a scalar field and $w^a$ are coordinates on $\Sigma$.

By the Einstein equations, the initial data on $\Sigma$ must satisfy the following constraints:
\begin{align}\label{constraint1}
&r_{\Sigma} + K_{\Sigma}^{2} - (K_{\Sigma})_{ab} (K_{\Sigma})^{ab} = 16 \pi G \, T_{\mu\nu} t^{\mu} t^{\nu}~,\\
& D^{a} (K_{\Sigma})_{a b} - D_{b} K_{\Sigma} = 8 \pi G\, T_{b\nu} t^{\nu}~, \label{constraint2}
\end{align}
where $D_{a}=P^{\mu}_{a}\nabla_{\mu}$ is the covariant derivative that $\Sigma$ inherits from $(\mathcal{M},g_{\mu\nu})$; $r_{\Sigma}$ is the Ricci scalar intrinsic to $\Sigma$; and $K_{\Sigma}$ is the trace of the extrinsic curvature: $K_{\Sigma} = (h_{\Sigma})^{ab}(K_{\Sigma})_{a b}$.

Let $\Sigma$ be a Cauchy slice of $\mathcal{M}$ containing $\mathcal{R}$ and smooth in a neighborhood of $\mathcal{R}$. The kink transform is then a map of the initial data on $\Sigma$ to a new initial data set, parametrized by a real number $s$ analogous to boost rapidity.  The transform acts as the identity on all data except for the extrinsic curvature, which is modified only at the location of the extremal surface $\mathcal{R}$, as follows:
\begin{align}\label{onlychange}
 (K_{\Sigma})_{ab}  \to (K_{\Sigma_s})_{ab} = (K_{\Sigma})_{ab} -\sinh{(2\pi s)}~ x_{a} x_{b} ~\delta(\mathcal{R})~.
\end{align}
Here $x^{a}$ is a unit norm vector field orthogonal to $\mathcal{R}$ and tangent to $\Sigma$, and we define
\begin{align}
\delta(\mathcal{R})\equiv \delta(x)~,
\end{align}
where $x$ is the Gaussian normal coordinate to $\mathcal{R}$ in $\Sigma$ ($\partial_{x} = x^{a}$). Thus, the only change in the initial data is in the component of the extrinsic curvature normal to $\mathcal{R}$. An equivalent transformation exists for initial choices of $\Sigma$ that are not smooth around $\mathcal{R}$ though the transformation rule will be more complicated than Eq.~\eqref{onlychange}. We will discuss this later in the section.

\begin{figure}
\center
        \includegraphics[width=1\textwidth]{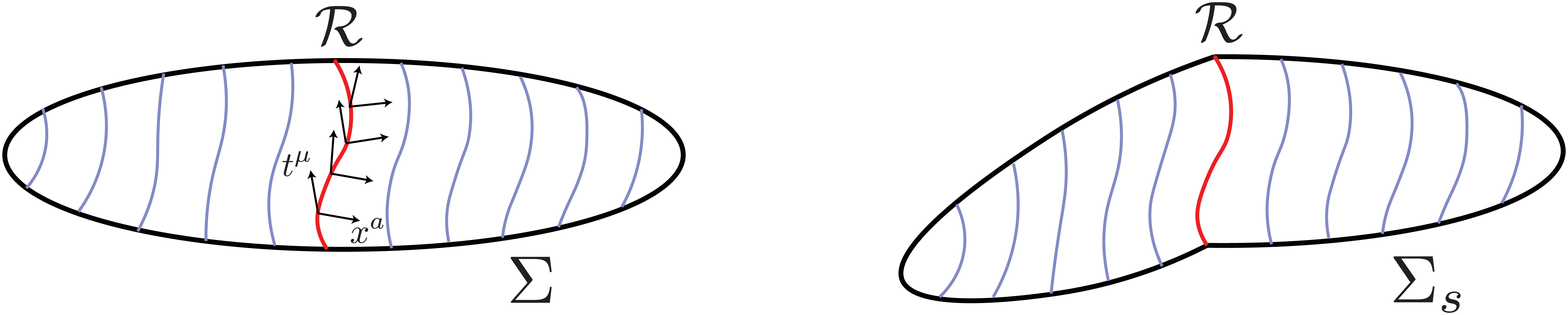}
        \caption{Kink transform. Left: a Cauchy surface $\Sigma$ of the original bulk $\mathcal{M}$. An extremal surface $\mathcal{R}$ is shown in red. The orthonormal vector fields $t^a$ and $x^{a}$ span the normal bundle to $\mathcal{R}$; $x^a$ is tangent to $\Sigma$. Right: The kink transformed Cauchy surface $\Sigma_{s}$. As an initial data set, $\Sigma_{s}$ differs from $\Sigma$ only in the extrinsic curvature at $\mathcal{R}$ through Eq.~\eqref{onlychange}. Equivalently, the kink transform is a relative boost in the normal bundle to $\mathcal{R}$, Eq.~\eqref{boostangle}.}
        \label{fig:kink2}
\end{figure}

Let $\Sigma_s$ be a time slice with this new initial data, as in Fig.~\ref{fig:kink2}, and let $\mathcal{M}_s$ be the Cauchy development of $\Sigma_s$. That is, $\mathcal{M}_s$ is the new spacetime resulting from the evolution of the kink-transformed initial data. Since the intrinsic metric of $\Sigma_s$ and $\Sigma$ are the same, they can be identified as $d$-manifolds with metric; the subscript $s$ merely reminds us of the different extrinsic data they carry. In particular the surface $\mathcal{R}$ can be so identified; thus $\mathcal{R}_s$ has the same intrinsic metric as $\mathcal{R}$. It also trivially has identical extrinsic data with respect to $\Sigma_s$. In fact, we will find below that like $\mathcal{R}$ in $\mathcal{M}$, $\mathcal{R}_s$ is an extremal surface in $\mathcal{M}_s$. However, the trace-free part of the extrinsic curvature of $\mathcal{R}_s$ in $\mathcal{M}_s$ may have discontinuities.

We will now show that the constraint equations hold on $\Sigma_s$; that is, the kink transform generates valid initial data. This need only be verified at $\mathcal{R}$ since the transform acts as the identity elsewhere. Here we will make essential use of the extremality of $\mathcal{R}$ in $\mathcal{M}$, which we express as follows.

The extrinsic curvature of $\mathcal{R}$ with respect to $\mathcal{M}$ has two independent components. Often these are chosen to be the two orthogonal null directions, but we find it useful to consider
\begin{align}
    &(B_{\mathcal{R}}^{(t)})_{ij} = P^{\mu}_{i}P^{\nu}_{j}\nabla_{(\mu} t_{\nu)}~,\\
    &(B_{\mathcal{R}}^{(x)})_{ij} = P^{\mu}_{i}P^{\nu}_{j}\nabla_{(\mu} x_{\nu)}~.
\end{align}
Here $i,j$ represent directions tangent to $\mathcal{R}$, and $P^{\mu}_{i}$ is the projector from $\mathcal{M}$ to $\mathcal{R}$. Extremality of $\mathcal{R}$ in $\mathcal{M}$ is the statement that the trace of each extrinsic curvature component vanishes:
\begin{align}
    &B_{\mathcal{R}}^{(t)}=(\gamma_{\mathcal{R}})^{ij}(B_{\mathcal{R}}^{(t)})_{ij} =0~,\\
    &B_{\mathcal{R}}^{(x)}=(\gamma_{\mathcal{R}})^{ij}(B_{\mathcal{R}}^{(x)})_{ij} =0~,
\end{align}
where $(\gamma_{\mathcal{R}})_{ij} = P^{a}_{i} P^{b}_{j} (h_{\Sigma})_{ab}$ is the intrinsic metric on $\mathcal{R}$.

Orthogonality of $t^\mu$ and $x^\mu$ implies that $P^\mu_i=P^a_i P^\mu_a$, and hence
\begin{equation}
  (B_{\mathcal{R}}^{(t)})_{ij} = P^{a}_{i}P^{b}_{j}(K_{\Sigma})_{ab}~.
  \label{BPP}
\end{equation}
%\begin{align}\label{R-extremality}
%(\gamma_{\mathcal{R}})^{ij}(K_{\mathcal{R}})_{ij} = 0~.
%\end{align}
Since $x^{a}$ is the unit norm orthogonal vector field at $\mathcal{R}$, the trace of $(K_{\Sigma})_{ab}$ at $\mathcal{R}$ can be written as:
\begin{align}
    \left. K_{\Sigma}\right|_{\mathcal{R}} = x^{a} x^{b}(K_{\Sigma})_{ab} + (\gamma_{\mathcal{R}})^{ij} (B_{\mathcal{R}}^{(t)})_{ij}= x^{a} x^{b}(K_{\Sigma})_{ab} ~.
\end{align}
A little algebra then implies
\begin{equation}
  (K_{\Sigma_s})^2-(K_{\Sigma_s})_{ab} (K_{\Sigma_s})^{ab} = (K_{\Sigma})^2-(K_{\Sigma})_{ab} (K_{\Sigma})^{ab}~.
\end{equation}
Moreover, we have $r_{\Sigma}=r_{\Sigma_s}$ since the two initial data slices have the same intrinsic metric. Thus Eq.~\eqref{constraint1} implies that the kink-transformed slice satisfies the scalar constraint equation:
\begin{align}\label{econstraint}
r_{\Sigma_s}+(K_{\Sigma_s})^2-(K_{\Sigma_s})_{ab} (K_{\Sigma_s})^{ab} =  16 \pi G T_{\mu\nu} t^{\mu} t^{\nu}~.
\end{align}

To check the vector constraint Eq.~\eqref{constraint2}, we separately consider the two cases of $b=x$ and $b=i$ where $i,j$ represent directions tangent to $\mathcal{R}$:
\begin{align}\label{momentC}
D_{a} (K_{\Sigma_s})^{a}_{x} - D_{x} K_{\Sigma_s} &= D_{a} (K_{\Sigma})^{a}_{x} - D_{x} K_{\Sigma} + B_{\mathcal{R}}^{(x)} \sinh{(2\pi s)} \delta (x)\nonumber \\
&= D_{a} (K_{\Sigma})^{a}_{x} - D_{x} K_{\Sigma} = 8\pi G\, T_{x\nu}t^\nu~, \\
D_{a} (K_{\Sigma_s})^{a}_{i} - D_{i} K_{\Sigma_s} &= D_{a} (K_{\Sigma})^{a}_{i} - D_{i} K_{\Sigma}= 8\pi G\, T_{i\nu}t^\nu~,
\end{align}
where the second line of the first equation follows from the extremality of $\mathcal{R}$.

%But we have $\Gamma^{x}_{ix} \sim g^{xx} \partial_{i}g_{xx} = 0$ %since $g_{xx}=1$ on $\mathcal{R}$ and $\Gamma^{i}_{ix} \sim g^{ij} %\partial_{x}g_{ij} = 0$ since $\mathcal{R}$ is extremal.

We conclude that the kink transform is a valid modification to the initial data. For both constraints to be satisfied after the kink, it was essential that $\mathcal{R}$ is an extremal surface. Thus the kink transform is only well-defined across an extremal surface. Note also that $\mathcal{R}_s\subset \Sigma_s$ is an extremal surface in $\mathcal{M}_s$. By Eq.~\eqref{BPP},
\begin{equation}
  (B_{\mathcal{R}_s}^{(t)})_{ij} = P^{a}_{i}P^{b}_{j}\left. (K_{\Sigma_s})_{ab}\right|_{\mathcal{R}_s} =
    (B_{\mathcal{R}}^{(t)})_{ij}  \implies B_{\mathcal{R}_s}^{(t)} =0 ~.
\end{equation}
In the second equality we used Eq.~\eqref{onlychange} as well as the fact that all relevant quantities are intrinsic to $\Sigma_{s}$, so $\mathcal{R}_s$ can be identified with $\mathcal{R}$. Moreover,
\begin{equation}
  (B_{\mathcal{R}_s}^{(x)})_{ij} = (B_{\mathcal{R}}^{(x)})_{ij} \implies B_{\mathcal{R}_s}^{(x)} =0~,
\end{equation}
since this quantity depends only on the intrinsic metrics of $\Sigma$ and $\Sigma_s$, which are identical.

\subsection{Properties}
\label{properties}

We will now establish important properties and an equivalent formulation of the kink transform.

\begin{figure}
\center
        \includegraphics[width=.6\textwidth]{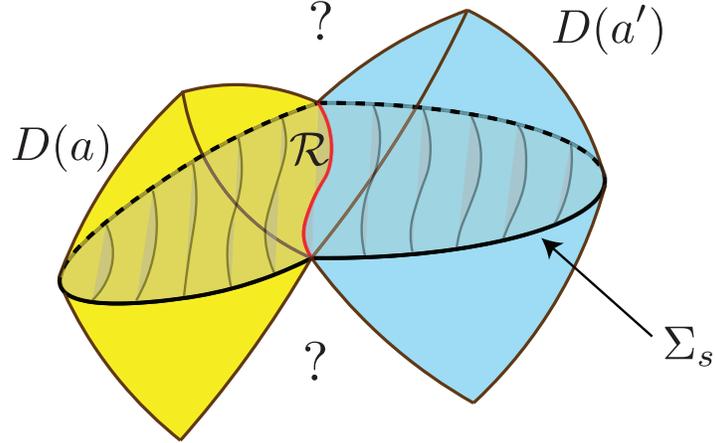}
        \caption{The kink-transformed spacetime $\mathcal{M}_s$ is generated by the Cauchy evolution of the kinked slice $\Sigma_{s}$. This reproduces the left and right entanglement wedges $D(a)$ and $D(a')$ of the original spacetime $\mathcal{M}$. The future and past of the extremal surface $\mathcal{R}$ are in general not related to the original spacetime.}
        \label{fig:KinkSolution2}
\end{figure}

Let us write $\Sigma$ as the disjoint union
\begin{equation}
    \Sigma = a'\cup \mathcal{R}\cup a~.
\end{equation}
The spacetime $\mathcal{M}$ contains $D(a)$ and $D(a')$ where $D(.)$ denotes the domain of dependence. The kink transformed slice $\Sigma_{s}$ contains regions $a$ and $a'$ with identical initial data, so $\mathcal{M}_{s}$ also contains $D(a)$ and $D(a')$. Because $\Sigma_s$ has different extrinsic curvature at $\mathcal{R}$, the two domains of dependence will be glued to each other differently in $\mathcal{M}_s$, so the full spacetime will differ from $\mathcal{M}$ in the future and past of $\mathcal{R}$. This is depicted in Fig.~\ref{fig:KinkSolution2}.

We will now derive an alternative formulation of the kink transform as a one-sided local Lorentz boost at $\mathcal{R}$. The unit vector field $t_{\Sigma_{s}}^{\mu}$ normal to $\Sigma_{s}$ is discontinuous at $\mathcal{R}$ due to the kink. Let
\begin{align}
    &({t_{\Sigma_{s}}^{\mu}})_{R} = \lim_{x \to 0^{+}} t_{\Sigma_{s}}^{\mu}~,\\
    &({t_{\Sigma_{s}}^{\mu}})_{L} = \lim_{x \to 0^{-}} t_{\Sigma_{s}}^{\mu}
\end{align}
be the left and right limits to $\mathcal{R}$. The metric of $\mathcal{M}_{s}$ is continuous since it arises from valid initial data on $\Sigma_{s}$. Therefore, the normal bundle of 1+1 dimensional normal spacetimes to points in $\mathcal{R}$ is well-defined. The above vector fields $({t_{\Sigma_{s}}^{\mu}})_{R}$ and $({t_{\Sigma_{s}}^{\mu}})_{L}$ belong to this normal bundle. Therefore at each point on $\mathcal{R}$, the two vectors can only differ by a Lorentz boost acting in 1+1 dimensional Minkowski space. The kink transform, Eq.~\eqref{onlychange}, implies:
\begin{align}\label{boostangle}
    ({t_{\Sigma_{s}}^{\mu}})_{R}= {(\Lambda_{2\pi s})}^{\mu}_{\nu} ({t_{\Sigma_{s}}^{\nu}})_{L}~,
\end{align}
where ${(\Lambda_{2\pi s})}^{\mu}_{\nu}$ is a Lorentz boost of rapidity $2\pi s$. In this sense, the kink transform resembles a local boost around $\mathcal{R}$. Alternatively, we can view Eq.~\eqref{boostangle} as the definition of the kink transform. This definition can be applied to Cauchy slices that are not smooth around $\mathcal{R}$, but it reduces to Eq.~\eqref{onlychange} in the smooth case.

This observation applies equally to any other vector field $\xi^\mu$ in the normal bundle to $\mathcal{R}$,  if $\xi^\mu$ has a smooth extension into $D(a')$ and $D(a)$ in $\mathcal{M}$. The norm of $\xi^\mu$ and its inner products with $({t_{\Sigma_{s}}^{\mu}})_{L}$ and $({t_{\Sigma_{s}}^{\mu}})_{R}$ are unchanged by the kink transform. Hence, in $\mathcal{M}_{s}$, the left and right limits of $\xi^\mu$ to $\mathcal{R}$ will satisfy
\begin{equation}\label{cboostangle}
        \xi^{\mu}_{R}= {(\Lambda_{2\pi s})}^{\mu}_{\nu}  \xi^{\nu}_{L}~.
\end{equation}

Now let $\Xi\supset \mathcal{R}$ be another Cauchy slice of $D(\Sigma)$. Since $\Xi$ contains $\mathcal{R}$, its timelike normal vector field $\xi^{\mu}$ (at $\mathcal{R}$) lies in the normal bundle to $\mathcal{R}$. We have shown that Eq.~\eqref{boostangle} is equivalent to the kink transform of $\Sigma$; that Eq.~\eqref{cboostangle} is equivalent to the kink transform of $\Xi$; and that Eq.~\eqref{boostangle} is equivalent to Eq.~\eqref{cboostangle}. Hence the kink transform of $\Sigma$ is equivalent to the kink transform of $\Xi$. In other words, the spacetime resulting from a kink transform about $\mathcal{R}$ does not depend on which Cauchy surface containing $\mathcal{R}$ we apply the kink transform to.

\begin{figure}[t]
\begin{center}
  \includegraphics[width=0.9\textwidth]{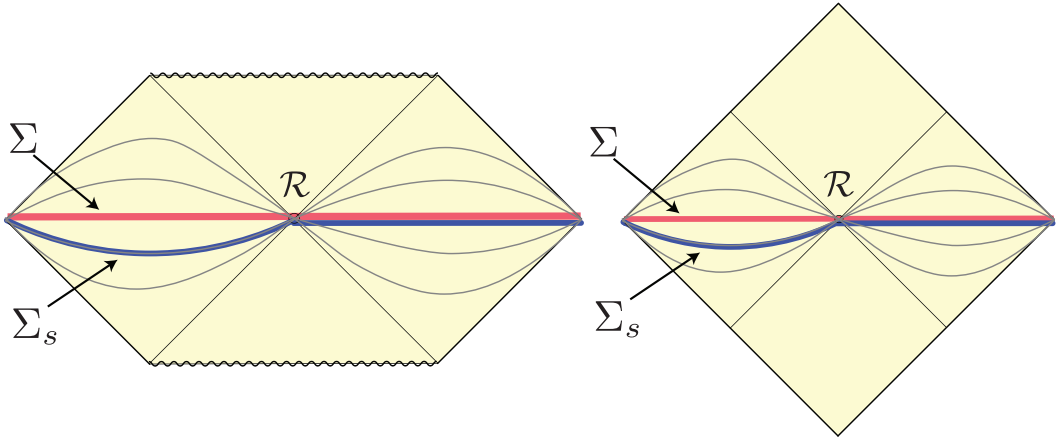}
\end{center}
\caption{Straight slices $\Sigma$ (red) in a maximally extended Schwarzschild (left) and Rindler (right) spacetime get mapped to kinked slices $\Sigma_s$ (blue) under the kink transform about $\mathcal{R}$.
}
\label{fig:kink}
\end{figure}

The kink transform (with $s\neq 0$) always generates physically inequivalent initial data. However $\mathcal{M}_s$ need not differ from $\mathcal{M}$. They will be the same if and only if $\Sigma_s$ is an initial data set in $\mathcal{M}$. There is an interesting special case where this holds for all values of $s$. Namely, suppose $\mathcal{M}$ has a Killing vector field that reduces to a boost in the normal bundle to $\mathcal{R}$. Then $\Sigma_s\subset \mathcal{M}$ (as a full initial data set), for all $s$. For example, the kink transform maps straight to kinked slices in the Rindler or maximally extended Schwarzschild spacetimes (see Fig.~\ref{fig:kink}).

\begin{figure}[t]
\begin{center}
  \includegraphics[scale=0.16]{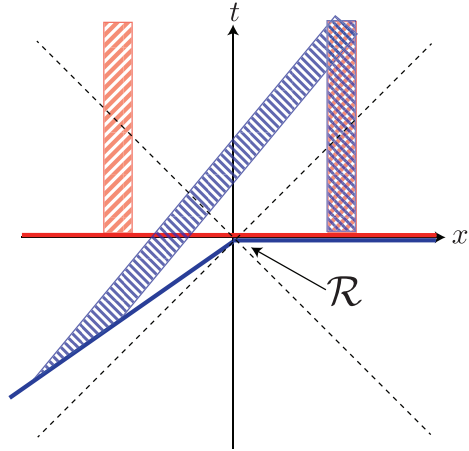}
\end{center}
\caption{On a fixed background with boost symmetry, the kink transform changes the initial data of the matter fields. In this example, $\mathcal{M}$ is Minkowski space with two balls relatively at rest (red).% at $x=\pm1$,$y=z=0$. The $t=0$ slice (red) becomes the slice (blue) kinked about the origin $\mathcal{R}$ under the kink transform.
The kink transform is still Minkowski space, but the balls collide in the future of $\mathcal{R}$ (blue). 
}
\label{fig:dust}
\end{figure}

We can also consider the kink transform of matter fields on a fixed background spacetime with the above symmetry. Geometrically, $\mathcal{M}=\mathcal{M}_s$ for all $s$,
%$\Sigma_s$ is embedded in $\mathcal{M}$ for all $s$. 
but the matter fields will differ in $\mathcal{M}_s$ by a one-sided action of the Killing vector field. For example, let $\mathcal{M}$ be Minkowski space, with two balls at rest at $x=\pm 1$, $y=z=0$ (see Fig.~\ref{fig:dust}); and let $\mathcal{R}$ given by $x=t=0$.  In the spacetime $\mathcal{M}_s$ obtained by a kink transform, the two balls will approach with velocity $\tanh 2\pi s$ and so will collide. The right and left Rindler wedge, $D(a)$ and $D(a')$, are separately preserved; the collision happens in the past or future of $\mathcal{R}$.

%{\color{red}(SKIP THE TOP PARAGRAPH WHEN READING THIS) The kink transform can also be applied in the presence of matter excitations on spacetimes with such Killing fields. Consider a field $\phi(x)$ in $\mathcal{M}$ arising from some initial data set on $\Sigma$. From this, we can generate a one-parameter family of solutions $\phi_{s}(x) = \phi(\Phi_{2\pi s}(\xi))$ where $\Phi_{(.)}(\xi)$ is the restriction of the Killing flow to $D(a')$. It is evident that this new solution results in equivalent initial data set on $\Sigma_{s}$, since $\Sigma_{s}$ is related to $\Sigma$ by the same one-sided flow. Therefore, the kink transform results in the same transformation of $\phi(x)$ as the action of a ``one-sided boost''.}

\section{Bulk Kink Transform = Boundary CC Flow}
\label{duality}

In this section, we will argue that the kink transform is the bulk dual of boundary CC flow. We will show that the kink transform satisfies two nontrivial necessary conditions. First, in Sec.~\ref{MLR}, we show that the left and right bulk region are the subregion duals to the left and right boundary region, respectively. In Sec.~\ref{shocks} we show that the bulk kink transform leads to precisely the stress tensor shock at the boundary generated by boundary CC flow, Eq.~\eqref{spacelikeshock}. (In Sec.~\ref{prediction} we will show that the kink transform predicts additional shocks in the CC flowed state, which have not been derived previously purely from QFT methods.)

\subsection{Matching Left and Right Reduced States}
\label{MLR}

The entanglement wedge of a boundary region $A$ in a (pure or mixed) state $\rho_A$,
\begin{equation}
    \text{EW}(\rho_A) = D[a(\rho_A)]
\end{equation}
is the domain of dependence of a bulk achronal region $a$ satisfying the following properties~\cite{Ryu:2006bv,Hubeny:2007xt,Faulkner:2013ana,Engelhardt:2014gca}:
\begin{enumerate}
    \item The topological boundary of $a$ (in the unphysical spacetime that includes the conformal boundary of AdS) is given by $\partial a = A\cup \mathcal{R}$.
    \item $S_{\rm gen}(a)$ is stationary under small deformations of $\mathcal{R}$.
    \item Among all regions that satisfy the previous criteria, EW$(\rho_A)$ is the one with the smallest $S_{\rm gen}(a)$.
\end{enumerate}
We neglect end-of-the-world branes in this discussion~\cite{Takayanagi:2011zk,Kourkoulou:2017zaj}. The generalized entropy is given by 
\begin{equation}
    S_{\rm gen} = \frac{\mbox{Area}(\mathcal{R})}{4G\hbar} + S(a) + \ldots~,
    \label{sgendef}
\end{equation}
where $S(a)$ is the von Neumann entropy of the region $a$ and the dots indicate subleading geometric terms. The entanglement wedge is also referred to as the Wheeler-DeWitt patch of $A$. 

There is significant evidence~\cite{Dong:2016eik,Hayden:2018khn} that EW$(\rho_A)$ represents the entire bulk dual to the boundary region $A$. That is, all bulk operators in EW$(\rho_A)$ have a representation in the algebra of operators $\cal{A}$ associated with $A$; and all simple correlation functions in $A$ can be computed from the bulk. In other words, the entanglement wedge appears to be the answer~\cite{Engelhardt:2014gca} to the question~\cite{Bousso:2012sj,Bousso:2012mh,Czech:2012bh,Hubeny:2012wa} of ``subregion duality.'' A bulk surface $\mathcal{R}$ is called quantum extremal (with respect to $A$ in the state $\rho$) if it satisfies the first two criteria, and quantum RT if it satisfies all three. When the von Neumann entropy term in Eq.~\eqref{sgendef} is neglected, $\mathcal{R}$ is called an extremal or RT surface, respectively. This will be the case everywhere in this paper except in Sec.~\ref{sec:quantum}.

We now specialize to the setting in which CC flow was considered in Sec.~\ref{CC}. Recall that the pure boundary state $\ket{\psi(\mathcal{C})}$ is given on a boundary slice $\mathcal{C}$ corresponding to $t=0$ in standard Minkowski coordinates; and that we regard $\mathcal{C}$ as the disjoint union of the left region $A'_0$ ($x<0$), with reduced state $\rho^\psi_{A_0}$; the cut $\partial A_0$ ($x=0$); and the right region $A_0$ ($x>0$), with reduced state $\rho^\psi_{A'_0}$. Let $a'_0$ and $a_0$ be arbitrary Cauchy surfaces of the associated entanglement wedges EW$(\rho^\psi_{A'_0})$ and EW$(\rho^\psi_{A_0})$. 

The entanglement wedges of non-overlapping regions are always disjoint, so
\begin{equation}
    \text{EW}(\rho^\psi_{A'_0})\cap \text{EW}(\rho^\psi_{A_0}) = \varnothing~.
\end{equation}
For the bipartition of a pure boundary state $\psi$, entanglement wedge complementarity holds:
\begin{equation}
a[\ket{\psi(\mathcal{C})}] = a'_0\cup \mathcal{R} \cup a_0~,
\label{cauchyparts}
\end{equation}
where $a[\ket{\psi(\mathcal{C})}]$ is a Cauchy surface of EW$(\ket{\psi(\mathcal{C})})$. In particular, the left and right entanglement wedge share the same HRT surface $\mathcal{R}$. 

Crucially, the classical initial data on $a[\ket{\psi(\mathcal{C})}]$ is almost completely determined by the data on $a'_0$ and $a_0$; however the data on $\mathcal{R}$ are not contained in $a'_0$ nor in $a_0$. In the semiclassical regime, the quantum state on $a[\ket{\psi(\mathcal{C})}]$ also includes global information (through its entanglement structure) that neither subregion contains on its own. Hence in general 
\begin{equation}
    \text{EW}(\ket{\psi(\mathcal{C})}) = D\left[\text{EW}(\rho^\psi_{A'_0})\cup \mathcal{R} \cup \text{EW}(\rho^\psi_{A_0})\right]
\end{equation}
is a proper superset of $\text{EW}(\rho^\psi_{A'_0})\cup \text{EW}(\rho^\psi_{A_0})$ that also includes some of the past and future of $\mathcal{R}$.

Now consider the CC-flowed state on the precursor slice $\ket{\psi_s(\mathcal{C}_s)}$. By Eqs.~\eqref{CCkink11} and \eqref{CCkink22}, we have
\begin{align}
    \text{EW}(\rho^{\psi_s}_{A'_s}) & = \text{EW}(\rho^\psi_{A'_0}) = D(a'_0)~,\\
    \text{EW}(\rho^{\psi_s}_{A_0}) & =  \text{EW}(\rho^\psi_{A_0})= D(a_0)~,
\end{align}
Since $\ket{\psi_s}$ is again a pure state, EW$[\ket{\psi_s(\mathcal{C}_s)}] = D\left(a[\ket{\psi_s({\mathcal{C}_s})}]\right)$ where
\begin{equation}
    a[\ket{\psi_s({\mathcal{C}_s})}] = a'_0\cup \mathcal{R} \cup a_0~.
    \label{newcauchy}
\end{equation}
We see that this initial data slice has the same intrinsic geometry as that of the original bulk dual. Indeed, by the remarks following Eq.~\eqref{cauchyparts}, the full classical initial data for the bulk dual to $\ket{\psi_s}$ will be identical on $a'_0\cup a_0$ and can only differ from the initial data for the original bulk at $\mathcal{R}$.

We pause here to note that a kink transform of $a[\ket{\psi(\mathcal{C})}]$ centered on $\mathcal{R}$ satisfies this necessary condition and hence becomes a candidate for $a[\ket{\psi_s({\mathcal{C}_s})}]$. However, this does not yet constrain the value of $s$. In order to go further, we would now like to show that a kink transform of $a[\ket{\psi(\mathcal{C})}]$ with parameter $s$ yields a bulk slice whose boundary is geometrically the precursor slice $\mathcal{C}_s$.

The bulk metric takes the asymptotic form \cite{Fefferman:2007rka}:\footnote{We set $\ell_{\text{AdS}} = 1$.}
\begin{equation}
    ds^2 = \frac{1}{z^2}\left[dz^2 + \eta_{AB} dx^A dx^B + O(z^d) \right]~,
    \label{dumbfg}
\end{equation}
where $\eta_{AB}$ is the metric of Minkowski space. Consider a stationary bulk surface $\mathcal{R}$ anchored on the boundary cut $u=v=0$. At leading order, $\mathcal{R}$ will reside at $u=v=0$ in the asymptotic bulk, in the above metric~\cite{Koeller:2015qmn}. (The first subleading term, which appears at order $z^d$, will be crucial in our derivation of the boundary stress tensor shock in Sec.~\ref{shocks}.) 

Let $\Sigma$ be a bulk surface that contains $\mathcal{R}$ and satisfies $t=0+O(z^d)$ in the metric of Eq.~\eqref{dumbfg}. Since the initial data on each side of $\mathcal{R}$ are separately preserved (see Sec.~\ref{properties}), Eq.~\eqref{boostangle} dictates that the kink transform $\Sigma_s$ of $\Sigma$ satisfies $t=0$ ($x>0$) and $t=x\tanh 2\pi s$ ($x<0$), again up to corrections of order $z^d$. %From Eq.~\eqref{dumbfg} it can be seen that $\Sigma_s$ satisfies Eq.~\eqref{onlychange} and so is the kink transform of $\Sigma$ along $\mathcal{R}$ at order $z^0$. (We ignore the matter data here since only the geometric data are relevant.)
The corrections all vanish at $z=0$, where $\Sigma$ is bounded by $\mathcal{C}$ and $\Sigma_s$ is bounded by $\mathcal{C}_s$ (see Eq.~\eqref{aps}). Recall also that the kink transform is slice-independent. Thus we have established that the kink transform of any Cauchy surface $a[\ket{\psi(\mathcal{C})}]$, by $s$ along $\mathcal{R}$, yields a Cauchy surface bounded by the precursor slice $\mathcal{C}_s$.

The above arguments establish that
\begin{equation}
    \text{EW}[\ket{\psi_s(\mathcal{C}_s)}] = D\left(a[\ket{\psi_s({\mathcal{C}_s})}]\right)~,
\end{equation}
where $a[\ket{\psi_s({\mathcal{C}_s})}]$ is given by Eq.~\eqref{newcauchy}. In words, the bulk dual of the CC-flowed boundary state is the Cauchy development of the kink-transform of a Cauchy slice containing the HRT surface $\mathcal{R}$. Note that the classical initial data on this Cauchy surface is fully determined by the initial data on $a'_0$ and $a_0$ inherited from the bulk dual of $\ket{\psi(\mathcal{C})}$, combined with the distributional geometric initial data consisting of the extrinsic curvature shock at $\mathcal{R}$. The full spacetime geometry will differ from EW$[\ket{\psi(\mathcal{C})}]$ because of the different gluing at $\mathcal{R}$.

% \begin{figure}\label{fig-Bulk}
% \center
%         \includegraphics[width=.4\textwidth]{Bulk}
%         \caption{}
% \end{figure}

\subsection{Matching Bulk and Boundary Shocks}
\label{shocks}

In Sec.~\ref{kink}, we gave a prescription for generating bulk geometries in AdS by inserting a kink on the Cauchy surface, at the HRT surface. With the standard holographic dictionary, the resulting geometry manifestly yields the correct behavior of one-sided boundary observables under CC flow. This was shown in the previous subsection. 

Another characteristic aspect of the CC flowed state $|\psi_s\rangle$ is the presence of a stress tensor shock at the cut (Sec.~\ref{stsc}), proportional to shape derivatives of the von Neumann entropy; see Eq.~\eqref{spacelikeshock}. We will now verify that this shock is reproduced by the kink transform in the bulk, upon applying the AdS/CFT dictionary. Notably, the shock is not localized to either wedge. Verifying kink/CC duality for this observable furnishes an independent, nontrivial check of our proposal.

We will now keep the first subleading term in the Fefferman-Graham expansion of the asymptotic bulk metric~\cite{Fefferman:2007rka,Koeller:2015qmn}:
\begin{align}
    ds^2 &= \frac{1}{z^2}\left(dz^2 + g_{AB}(x,z)dx^A dx^B \right)~,\\ 
    g_{AB}(x,z) &= \eta_{AB} +z^d \frac{16\pi G}{ d }\langle T_{AB}\rangle + o(z^d)\label{gexp}~,
\end{align}
where indices $A,B,\ldots$ correspond to directions along $z = \text{const.}$ surfaces.

The location of the RT surface $\mathcal{R}$ in the bulk can be described by a collection of $(d-1)$ embedding functions
\begin{equation}
    X^{\mu}(y,z) = (z,X^A(y,z))~,
\end{equation}
where $(y,z)$ are intrinsic coordinates on $\mathcal{R}$. The expansion in $z$ takes the simple form 
\begin{equation}
\label{xexp}
   X^A(y,z)) = z^d X^A_{(d)} + o(z^d) ~,
\end{equation}
because the boundary anchor is the flat cut $u=v=0$ of the Rindler horizon~\cite{Koeller:2015qmn}. Stationarity of $\mathcal{R}$ can be shown to imply~\cite{Koeller:2015qmn}
\begin{equation}
\label{holoS}
   X^A_{(d)} = -\frac{4 G}{d} \left. \frac{\delta S}{\delta X^A}\right|_{\mathcal{R}}~.
\end{equation}

We consider a bulk Cauchy slice $\Sigma\supset\mathcal{R}$ for which $\partial \Sigma$ corresponds to the $t = 0$ slice on the boundary. Since the subleading terms in Eqs.~\eqref{gexp} and \eqref{xexp} start at $z^d$, we are free to choose $\Sigma$ so that it is given by 
\begin{equation} 
t = z^d \varsigma(x)+o(z^d)~,
\label{vars1}
\end{equation}
Recall that the vector fields $t^{\mu}$ and $x^{\mu}$ are defined to be orthogonal to $\mathcal{R}$, and respectively orthogonal and tangent to $\Sigma_s$. In FG coordinates one finds:
\begin{align}
    t^{A} &= z \left(t^{A}_{(0)} + z^d t^{A}_{(d)} + o(z^d)\right) \label{tacomp}~, \\
    x^{A} &= z \left(x^{A}_{(0)} + z^d x^{A}_{(d)} + o(z^d)\right) \label{xacomp}~, \\
    t^z &= z\left(z^{d-1} t^{z}_{(d-1)} + o(z^{d-1})\right) \label{tzcomp}~, \\
    x^z &= z\left(z^{d-1} x^z_{(d-1)} + o(z^{d-1})\right)~. \label{xzcomp}~.
\end{align}
The overall factor of $z$ is due to normalization. Note that $t^\mu_{(0)}$ is a coordinate vector field but in general, $t^\mu$ is not. Individual coordinate components of vectors and tensors are defined by contractions with $t^{\mu}_{(0)}$ and $x^{\mu}_{(0)}$ respectively, for example $t^t\equiv t_\mu t^\mu_{(0)}$.

We now consider a contraction of the extrinsic curvature tensor on $\Sigma$,
\begin{align}
  (K_\Sigma)_{ab}x^b = P^{\mu}_a x^\nu \nabla_{(\mu}t_{{\nu})}~. \label{shockextrinsic}
\end{align}
We would like to further project the $a$ index onto the $z$ direction. Deep in the bulk the $z$ direction does not lie entirely in $\Sigma$. However, note that $g_{\mu z}t^{\mu} \rightarrow 0$ in the limit $z\rightarrow 0$ due to Eq.~\eqref{tzcomp}. Therefore, at leading order in $z$, the $z$ direction does lie entirely in $\Sigma$; moreover, $P^{\mu}_z \rightarrow \delta^{\mu}_z$ as $z\rightarrow 0$. We will only be interested in evaluating Eq.~\eqref{shockextrinsic} at leading order in $z$ so we may freely set $a = z$, which yields:\footnote{In $d> 2$ the terms involving $x^z t^z$ will be higher order, by Eqs.~\eqref{tzcomp} and \eqref{xzcomp}, and need not be included. Since they cancel out either way, we include them here to avoid an explicit case distinction.}
\begin{align}
 (K_\Sigma)_{z\nu}x^\nu - x^\nu \partial_{(z}t_{{\nu})} 
 &= x^{\nu} t_{\gamma}\, \Gamma^{\gamma}_{\nu z}\\
 &= z^2 \Gamma_{txz} + z x^t \Gamma_{ttz} 
    + z t^x\Gamma_{xxz}+ x^z t^z \Gamma_{zzz}+o(z^{d-1})\\
 &= \frac{(d-2)}{2}z^{d-1}\frac{16\pi G}{d}\langle T_{tx}\rangle
 - z^{-3}t^z x^z - z^{d-1}(t^x_{(d)}-x^t_{(d)})
 +o(z^{d-1})~.
\end{align}
The condition $x_{\mu}t^{\mu}=0$ implies that
\begin{align}
    z^{d-1}\frac{16\pi G}{ d }\langle T_{tx}\rangle + z^{-3}x^z t^z + z^{d-1}(t^x_{(d)} - x^t_{(d)}) +o(z^{d-1}) = 0~.
\end{align}
Hence we find 
\begin{equation}
    (K_{\Sigma})_{z\nu}x^\nu - x^\nu \partial_{(z}t_{{\nu})} = z^{d-1}\, 8\pi G\,\langle T_{tx}\rangle+o(z^{d-1})\label{kt}~.
\end{equation}

We now apply the kink transform to  $\Sigma$ (viewed as an initial data set). This yields a new initial data set on a slice $\Sigma_s$ in a new spacetime $\mathcal{M}_s$. We again expand in Fefferman-Graham coordinates:
\begin{align}
    ds^2 &= \frac{1}{z^2}\left(dz^2 + \tilde{g}_{AB}(\tilde x,z) d\tilde x^A d\tilde x^B \right)~,\\ 
    \tilde{g}_{AB}(x,z) &= \tilde \eta_{AB} +z^d \frac{16\pi G}{ d }\langle \tilde T_{AB}\rangle + o(z^d)\label{tildegexp}~.
\end{align}
Here $\tilde \eta_{AB}$ is still Minkowski space; any change in the bulk geometry will be encoded in the subleading term. 

The notation $\tilde \eta_{AB}$ indicates that we will be using the specific coordinates in which the metric of $d$-dimensional Minkowski space takes the nonstandard form given by Eq.~\eqref{tildeeta}. This has the advantage that the {\em coordinate} form of all vectors, tensors, and embedding equations in $D(a')\cup \mathcal{R}\cup D(a)$ will be unchanged by the kink transform, if we use standard Cartesian coordinates before the transform and the tilde coordinates afterwards. 

For example, the invariance of the left and right bulk domains of dependence under the kink transform implies that $\Sigma_s$ is given by
\begin{equation}
    \tilde t = z^d \varsigma(\tilde x)+o(z^d)~,
\end{equation}
with the {\em same} $\varsigma$ that appeared in Eq.~\eqref{vars1}. (In fact, this extends to at all orders in $z$.) As already shown in the previous subsection, $\partial \Sigma_s$ lies at $\tilde{t}=0$, $z=0$. 

As another example, the coordinate components of the unit normal vector to $\Sigma_s$ in $\mathcal{M}_s$, $\tilde t^\mu$, will be the same as the components of the normal vector to $\Sigma$ in $\mathcal{M}$, $t^\mu$, and therefore
\begin{equation}
    \left. \partial_{(z}t_{{\nu})}\right|_{\mathcal{M}} = 
    \left. \partial_{(z}\tilde t_{{\nu})}\right|_{\mathcal{M}_s}~.
    \label{partials}
\end{equation}
Below we will use the convention that any quantity with a tilde is evaluated in $\mathcal{M}_s$, in the coordinates of Eq.~\eqref{tildegexp}. Any quantity without a tilde is evaluated in $\mathcal{M}$, in the coordinates of Eq.~\eqref{gexp}. The only exception is the extrinsic curvature tensor, where the corresponding distinction is indicated by the subscript $\Sigma_s$ or $\Sigma$, for consistency with Sec.~\ref{kink}.

We now consider the extrinsic curvature of $\Sigma_s$. A calculation analogous to the derivation of Eq.~\eqref{kt} implies
\begin{equation}
    (K_{\Sigma_s})_{z\nu}\tilde x^\nu - \tilde x^\nu \partial_{(z}\tilde t_{{\nu})} = z^{d-1}\, 8\pi G\,\langle \tilde T_{\tilde t\tilde x}\rangle+o(z^{d-1})\label{tkt}~.
\end{equation}
From Eqs.~\eqref{kt} and \eqref{partials} we find
\begin{align}
    z^{d-1} \langle \tilde T_{\tilde t\tilde x}\rangle &=  z^{d-1} \langle T_{tx}\rangle + \frac{1}{8\pi G}\left[(K_{\Sigma_s})_{z\nu} - (K_{\Sigma})_{z\nu}\right] x^\nu+o(z^{d-1})~,\\
    &= z^{d-1} \langle T_{tx}\rangle -\frac{\sinh{(2\pi s)}}{8\pi G}\delta(\mathcal{R})x_{z}+o(z^{d-1}) ~.
\end{align} 
In the first equality, we used the fact that $x^\mu$ and $\tilde x^\mu$ can be identified as vector fields, and the extrinsic curvature tensors can be compared, in the submanifold $\Sigma=\Sigma_s$. The second equality follows from the definition of the kink transform, Eq.~\eqref{onlychange}.

By Eq.~\eqref{xacomp}, $\delta(\mathcal{R})=\delta(z^{-1}\tilde{x}) = z\delta(\tilde{x})$. The condition $x_{\mu} \partial_z X^{\mu} = 0$ yields 
\begin{align}
    x_z = -d\, z^{d-2}\widetilde{X}_{(d)}+o(z^{d-2})~,
\end{align}
where $\widetilde{X}_{(d)}$ is the $A = \tilde{x}$ component of $X^A_{(d)}$. Taking $z\to 0$ and using Eq.~\eqref{holoS}, we thus find
\begin{equation}
    \langle \tilde T_{\tilde t\tilde x}\rangle = \langle T_{tx}\rangle + \frac{1}{2\pi} \sinh(2\pi s)\frac{\delta S}{\delta \widetilde{X}}\Big \lvert_{\widetilde{X}=0}\delta(\tilde{x}) ~,
    \label{last}
\end{equation}
which agrees precisely with Eq.~\eqref{spacelikeshock}.

Note that this derivation applies to any boosted coordinate system $(\check t, \check x)$ as well. Linear combinations of Eq.~\eqref{last} with its boosted version reproduces both the $T_{\tilde u\tilde u}$ shock of Eq.~\eqref{Ttutu} and the $T_{\tilde v\tilde v}$ shock of  Eq.~\eqref{shocktransform} holographically.

\section{Predictions}\label{prediction}

Having found nontrivial evidence for kink transform/CC flow duality, we now change our viewpoint and assume the duality. In this section, we will derive a \emph{novel} property of CC flow from the kink transform: a shock in the $\langle T_{xx}\rangle$ component of the stress tensor in the CC flowed state. We do not yet know of a way to derive this directly in the quantum field theory, so this result demonstrates the utility of the kink transform in extracting nontrivial properties of CC flow. We further argue that $\langle T_{xx}\rangle$ and $\langle T_{tx}\rangle$ constitute all of the independent, nonzero stress tensor shocks in the CC flowed state. 

Our holographic derivation only depends on near boundary behavior, and the value of the shock takes a universal form similar to Eq.~\eqref{last}. Thus, we expect that the properties we find in holographic CC flow hold in non-holographic QFTs as well. 

To derive the $\langle T_{xx}\rangle$ shock, we use the Gauss-Codazzi relation~\cite{Gourgoulhon:2007ue}
\begin{align}
    P^{\mu}_a P^{\nu}_b P^{\alpha}_c P^{\beta}_d R_{\mu \nu \alpha \beta} = K_{ac}K_{bd} - K_{bc}K_{ad} + r_{abcd}~, \label{Gauss}
\end{align}
where $r_{abcd}$ is the intrinsic Riemann tensor of $\Sigma$. It is important to note that this relation is purely intrinsic to $\Sigma$. Since $\Sigma=\Sigma_s$ as submanifolds, we can not only evaluate Eq.~\eqref{Gauss} in both $\mathcal{M}$ and $\mathcal{M}_s$ but also meaningfully subtract the two. We emphasize that the following calculation is only nontrivial in $d > 2$ (in $d = 2$, the Gauss-Codazzi relation is trivial). We comment on $d = 2$ at the end.  

First we evaluate Eq.~\eqref{Gauss} in $\mathcal{M}$. We will only be interested in evaluating it to leading order in $z$ in the Fefferman-Graham expansion. As argued in Sec.~\ref{shocks}, when working at leading order we can freely set $a = c = z$. We then compute the following at leading order in $z$:
\begin{align}
    R_{z x z  x} = K_{zz}K_{x x} - (K_{x z})^2 + r_{z x z x}~.
\end{align}

We start by computing $K_{zz}$. First we note that  $\Gamma_{zz}^{\alpha}t_{\alpha} = 0$ identically. Therefore, 
\begin{align}
    K_{zz} = \partial_z t_z = 4G (d-2)z^{d-3}\frac{\delta S}{\delta T}\Big \lvert_{\mathcal{R}}+o(z^{d-3})~ .
\end{align}
We have made use of 
\begin{align}
    t_{z} = 4G z^{d-2}\frac{\delta S}{\delta T}\Big \lvert_{\mathcal{R}}+o(z^{d-2})~,
\end{align}
which follows from $t_{\mu}\partial_z X^{\mu} = 0$. 

Next we compute
\begin{align}
    R_{zxzx} = \partial_x \Gamma^x_{zz} - \partial_z \Gamma^x_{xz}+ \Gamma^x_{x\mu}\Gamma^{\mu}_{zz}- \Gamma^{x}_{z\mu}\Gamma^{\mu}_{xz}~.
\end{align}
One finds 
\begin{align}
    \partial_z \Gamma^x_{xz} &= \frac{1}{2}(d-2)(d-1)z^{d-2}\frac{16\pi G}{d}\langle T_{xx}\rangle + o(z^{d-2})~, \\ 
    \Gamma^{x}_{xz}\Gamma^{z}_{zz} &= -\frac{1}{2}(d-2)z^{d-2}\frac{16\pi G}{d}\langle T_{xx}\rangle + o(z^{d-2})~, 
\end{align}
with all other terms either subleading in $z$ or identically vanishing, and hence
\begin{align}
    R_{zxzx} = -8\pi G(d-2)z^{d-2}\langle T_{xx}\rangle+o(z^{d-2})~.
\end{align}

Putting all this together, we have 
\begin{align}
    -8\pi G(d-2)z^{d-2}\langle T_{xx}\rangle= 4G z^{d-3}\frac{\delta S}{\delta \widetilde {T}}\Big \lvert_{\mathcal{R}}K_{xx} - (K_{xz})^2 + r_{zxzx} + o(z^{d-2})~. \label{gausseval}
\end{align}
The analogous relation evaluated in $\mathcal{M}_s$ reads 
\begin{align} \label{gaussevalaf}
    -8\pi G(d-2)z^{d-2}\langle \tilde T_{\tilde x \tilde x}\rangle= 4G z^{d-3}\frac{\delta S}{\delta \widetilde{T}}\Big \lvert_{\mathcal{R}}\tilde K_{\tilde x \tilde x} - (\tilde K_{\tilde x z})^2 + \tilde r_{z\tilde xz \tilde x} + o(z^{d-2})~,
\end{align}
where we have made use of Eq.~\eqref{partials} to set $K_{zz} = \tilde K_{zz}$. We can now subtract these two relations. First note that $\tilde r_{abcd} = r_{abcd}$ since it is purely intrinsic to $\Sigma$. Next, recall from the definition of the kink transform Eq.~\eqref{onlychange} that 
\begin{align}
    \tilde{K}_{\tilde x \tilde x} - K_{xx} = -z \sinh(2\pi s)\delta(\tilde{x}) ~.
\end{align}
Lastly, it is easy to check that $K_{xz}\sim o(z^{d-2})$ hence its contribution to Eq.~\eqref{gausseval} is subleading, and similarly for Eq.~\eqref{gaussevalaf}. Thus, subtracting Eq.~\eqref{gaussevalaf} from Eq.~\eqref{gausseval} yields 
\begin{align}
    \langle \tilde T_{\tilde x \tilde x}\rangle - \langle T_{xx}\rangle = \frac{1}{2\pi}\sinh(2\pi s)\frac{\delta S}{\delta \widetilde T}\Big \lvert_{\widetilde{X}=0}\delta(\tilde{x}) ~. \label{newshock}
\end{align}

The above calculation only works in $d > 2$. In $d = 2$, since the boundary theory is a CFT, tracelessness of the boundary stress tensor further implies that $\langle T_{tx}\rangle$ is the \emph{only} independent component of the stress tensor shock so there is no need for a calculation analogous to the one above. We expect that this argument is robust under relevant deformations of the CFT since the shock is highly localized and should universally depend only on the UV fixed point.

Together with the $\langle T_{tx}\rangle$ shock we reproduced in the previous section, and using Lorentz invariance of the boundary, this result determines the transformation of the the stress tensor contracted with any pair of linear combinations of $t$ and $x$, such as $\braket{T_{tt}}$. This linear space contains all of the independent nonvanishing components of the shock. To see this, note that
%\begin{comment}
%Finally, we have the following:  
\begin{align}
    &x^{\nu}\left(\nabla_{\nu}\tilde y_{\mu} - \nabla_{\nu}y_{\mu} \right) = 0~, \label{noshockone} \\
    &y^{\nu}\left(\nabla_{\nu}\tilde y_{\mu} - \nabla_{\nu}y_{\nu} \right)\label{noshocktwo} = 0~, \\
    &y^{\nu}\left( \nabla_{\nu}\tilde t_{\mu } - \nabla_{\nu}t_{\mu } \right) = 0~, \label{noshockthree}
\end{align}
where $y^{\mu} = P^{\mu}_i y^i$ for any vector field $y^i$ in the tangent bundle of $\mathcal{R}$. Eqs.~\eqref{noshockone} and \eqref{noshocktwo} follow trivially from the fact that the prescription Eq.~\eqref{cboostangle} only introduces a discontinuity in vector fields in the normal bundle of $\mathcal{R}$, while Eq.~\eqref{noshockthree} simply follows from Eq.~\eqref{onlychange}. Evaluating the $\mu=z$ components in the same way as in Sec.~\ref{shocks}, we find, 
\begin{align}
    &\langle \tilde{T}_{\tilde\mu \tilde y}\rangle - \langle T_{\mu y}\rangle = 0~.
\end{align}
%\end{comment}

For $s\to\infty$, the shocks derived in the previous two sections agree with those found to be required for the existence of certain coarse grained bulk states in Ref.~\cite{Bousso:2019dxk}. In that work, the cut was allowed to be a wiggly or flat cut of a bifurcate horizon such as a Rindler horizon, and the state could belong to any quantum field theory. Interpolation of these results suggests that the shocks we have derived here generalize to the case of CC flow for a wiggly cut of the Rindler horizon, in general QFTs with a conformal fixed point.

\section{Discussion}
\label{discussion}

\subsection{Relation to JLMS and One Sided Modular Flow}\label{sec:JLMS}

%Modular flow has played a crucial role in our understanding of AdS/CFT and bulk reconstruction \cite{Jafferis:2014lza,Jafferis:2015del,Faulkner:2017vdd,Chen:2019iro}.
The bulk dual of one-sided modular flow~\cite{Jafferis:2014lza,Faulkner:2018faa} resembles the kink transform. CC flow yields a well defined state, however, whereas a one sided modular flowed state is singular in QFT.
Correspondingly, the kink transform defined here yields a smooth bulk solution whereas the version implicitly defined in Ref.~\cite{Faulkner:2018faa} results in a singular spacetime (see also Ref.~\cite{Ceyhan:2018zfg}, footnote 4). We will now explain this in detail.
%, and we will show how our construction implements a nonperturbative generalization of the JLMS relation \cite{Jafferis:2015del}.

\begin{figure}[t]
\begin{center}
  \includegraphics[scale=0.2]{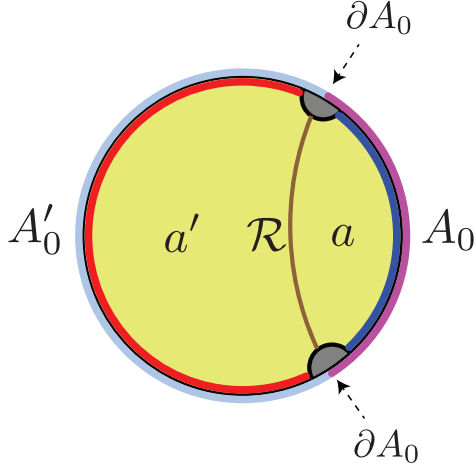} 
\end{center}
\caption{A boundary subregion $A_0$ (pink) has a quantum extremal surface denoted $\mathcal{R}$ (brown) and an entanglement wedge denoted $a$.
The complementary region $A'_0$ (light blue) has the entanglement wedge $a'$. CC flow generates valid states, but one-sided modular flow is only defined with a UV cutoff. For example, one can consider regulated subregions $A^{(\epsilon)}$ (deep blue) and $A'^{(\epsilon)}$ (red). In the bulk, this amounts to excising an infrared region (gray) from the joint entanglement wedge (yellow).}
\label{fig:JLMS}
\end{figure}

Consider a boundary region $A_0$ with reduced state $\rho_{A_0}$, dual to a semi-classical state $\rho_a$ in the bulk entanglement wedge $a$ associated to $A_0$ as seen in Fig.~\ref{fig:JLMS}.
We denote by $K_A = -\log \rho_A$ and $K_a= -\log \rho_a$ the boundary and bulk modular Hamiltonians, respectively. The JLMS result \cite{Jafferis:2015del} states that 
\begin{align}\label{JLMS}
    \hat{K}_{A_0} &= \frac{\hat{A}[\mathcal{R}]}{4G}+ \hat{K}_{a}~,
\end{align}
where $\hat{A}[\mathcal{R}]$ is the area operator that formally evaluates the area of the quantum extremal surface $\mathcal{R}$ \cite{Engelhardt:2014gca}.

Suppose now that $A_0$ has a nonempty boundary $\partial A_0$. Then there is an interesting asymmetry in Eq.~\eqref{JLMS}. The one-sided boundary modular operator appearing on the left hand side is well-defined only with a UV cutoff. On the other hand, at least the leading (area) term in the bulk modular operator on the right hand side has a well-defined action. Let us discuss each side in turn.

In Einstein gravity, the area operator $\hat A$ is the generator of one-sided boosts. To see this, let us restrict the gravitational phase space to the bulk region $D(\Sigma_{\epsilon})$. 
There exists a (non-unique) vector field $\xi^a$ in $D(a')\cup \mathcal{R}$ such that $\xi^a$ generates an infinitesimal one-sided boost at $\mathcal{R}$ \cite{Lashkari_2016,Donnelly_2016}. This boost can be quantified by a parameter $s$ in the normal bundle to $\mathcal{R}$, as described in Sec.~\ref{properties}. The area functional $A[\mathcal{R}]/4G$ is the Noether charge at $\mathcal{R}$ associated to $\xi^a$, given by the expectation value of the area operator in the semi-classical bulk state:
\begin{equation}
    A[\mathcal{R}] = \langle \hat{A}[\mathcal{R}]\rangle~.
\end{equation}

Each point in the gravitational phase space can be specified by the metric in $D(a')$, the metric in $D(a)$, and the boost angle $s$ at $\mathcal{R}$ with which the two domains of dependence are glued together \cite{Faulkner:2018faa, Donnelly_2016, Carlip_1995, Dong:2018seb}. The action of 
\begin{equation}
    \langle e^{2\pi i s\hat{A}[\mathcal{R}]/4G}\rangle
    \label{areaaction}
\end{equation} 
on points in the gravitational phase space is to simply shift the conjugate variable, {\em i.e.}, the relative boost angle between the left and right domains of dependence, by $s$. 
Note that the metrics in the left and right domains of dependence are unchanged since the area functional acts purely on the phase space data at $\mathcal{R}$. This is the classical analogue of the statement that the area operator is in the center of the algebras of the domains of dependence \cite{Harlow:2016vwg}. Comparing with Sec.~\ref{properties}, we see that this action is equivalent to the kink transform of $\Sigma_{\epsilon}$ about $\mathcal{R}$ by $s$. We stress that this action is well-defined even if $\mathcal{R}$ extends all the way out to the conformal boundary, {\em i.e.}, in the far ultra-violet from the boundary perspective.

We turn to the right hand side of Eq.~\eqref{JLMS}, still assuming that $A_0$ has a nonempty boundary $\partial A$. Since the algebra of a QFT subregion $A_0$ is a Type-III$_1$ von Neumann algebra, the Hilbert space does not factorize across $\partial A_0$~\cite{Witten:2018zxz}. A reduced density matrix $\rho_{A_0}$, and hence $\hat{K}_{A_0}$, do not exist. Physically, the action of $\hat{K}_{A_0}$ on a fixed boundary time slice would break the vacuum entanglement of arbitrarily short wavelength modes across $\partial A_0$; this would create infinite energy.

Therefore, any discussion of $\hat{K}_{A_0}$ requires introducing a UV regulator. Consider the regulated subregions $A_0^{(\epsilon)}$ and $A_0^{\prime(\epsilon)}$ shown in Fig.~\ref{fig:JLMS}. The split property in algebraic QFT~\cite{Witten:2018zxz,Dutta:2019gen} guarantees the existence of a (non-unique) Type-I von Neumann algebra $\mathcal{N}$ nested between the algebras of subregion $A_0^{(\epsilon)}$ and the complementary algebra of $A_0^{\prime(\epsilon)}$, {\em i.e.},
\begin{align}
    \mathcal{A}_0^{(\epsilon)} \subset \mathcal{N} \subset \left(\mathcal{A}_0^{\prime (\epsilon)}\right)'~.
\end{align}
With this prescription, one can define a regulated version of the reduced density matrix $\rho_A$ by using the Type-I factor $\mathcal{N}$ \cite{Doplicher:1984zz}.
%The bulk entanglement wedge of their union $A_{\epsilon}\cup A'_{\epsilon}$ is the domain of dependence of the partial Cauchy slice $\Sigma_{\epsilon}$ shaded yellow in Fig.~\ref{fig:JLMS}. 
It has been suggested that there exists an $\mathcal{N}$ consistent with the geometric cutoff shown in Fig.~\ref{fig:JLMS}~\cite{Takayanagi:2017knl,Dutta:2019gen}: the quantum extremal surface $\mathcal{R}$ in the bulk is regulated by a cutoff brane $B$ demarcating the entanglement wedge of the subregion $A_0^{(\epsilon)}\cup A_0^{\prime (\epsilon)}$. The regulated area operator $\hat{A}[\mathcal{R}]/4G$ is well defined once boundary conditions on $B$ are specified. %We do not expect our discussion to depend sensitively on the choice of $\mathcal{N}$ or the dynamics of the cutoff brane, since we will restrict to $D(\Sigma_{\epsilon})$.

Let us now specialize to the case for which we have conjectured kink transform/CC-flow duality: the boundary slice $\mathcal{C}=A_0\cup A'_0$ is a Cauchy surface of Minkowski space, and $\partial A_0$ is the flat cut $u=v=0$ of the Rindler horizon. We have just argued that the kink transformation is generated by the area operator through Eq.~\eqref{areaaction}. By Eq.~\eqref{JLMS}, the boundary dual of this action should be one-sided modular flow, not CC flow. By Eqs.~\eqref{ccflow} and \eqref{eq:CCrho}, these are manifestly different operations. Indeed, unlike one-sided modular flow, Connes cocycle flow yields a well-defined boundary state for all $s$, without any UV divergence at the cut $\partial A_0$: $\ket{\psi(\mathcal{C})}\to\ket{\psi_s(\mathcal{C})}$. 

In fact there is no contradiction. For both modular flow and CC flow on the boundary, a bulk-dual Cauchy surface $\Sigma_s$ is generated by the kink transform. The difference is in how $\Sigma_s$ is glued back to the boundary.

For modular flow, $\Sigma_s$ is glued back to the original slice $\mathcal{C}$. Generically, this would violate the asymptotically AdS boundary conditions, necessitating a regulator such as the excision of the grey asymptotic region in Fig.~\ref{fig:JLMS} and interpolation by a brane $B$. The boundary dual is an appropriately regulated modular flowed state with energy concentrated near the cut $\partial A_0$. This construction is possible even if $\partial A_0$ is not a flat plane, but the regulator is ambiguous and cannot be removed.\footnote{There is evidence that a code subspace can be defined with an appropriate regulator such that one-sided modular flow keeps the state within the code subspace \cite{Dong:2018seb,Akers:2018fow,Dong:2019piw}.}

For CC flow, $\Sigma_s$ is glued to the precursor slice $\mathcal{C}_s$ as discussed in Sec.~\ref{kink}. This yields $\ket{\psi_s(\mathcal{C}_s)}$. Time evolution on the boundary yields $\ket{\psi_s(\mathcal{C})}$, the CC-flowed state on the original slice $\mathcal{C}$.  

On the boundary, we can use the one-sided modular operator in two ways. As a map between states on $\mathcal{C}$~\cite{Jafferis:2015del,Faulkner:2017vdd} it requires a UV regulator. As a map that takes a state on $\mathcal{C}$ to a state on the precursor slice $\mathcal{C}_s$, $\ket{\psi(\mathcal{C})}\to\ket{\psi_s(\mathcal{C}_s)}$, it is equivalent to CC flow on $\mathcal{C}$ by Eqs.~\eqref{ccdo} and \eqref{precursor-rho}. This is a more natural choice due to its UV-finiteness. But it is available only if the vacuum modular operator for cut $\partial A$ is geometric, so that the precursor slice is well-defined.

\subsection{Quantum Corrections}
\label{sec:quantum}

It is natural to include semiclassical bulk corrections to all orders in $G$ to our proposal. The natural guess would be to perform the kink transform operation about the quantum extremal surface along with a CC flow for the bulk state.
In general, it is difficult to describe this procedure within EFT. In states far from the vacuum, the background spacetime changes under the kink transform, and it is unclear how to map states from one spacetime to another. However, we will find some evidence that suggests that the bulk operation relating the two states is a generalized version of CC flow in curved spacetime.

To see this, note that the quantum extremal surface $\mathcal{R}$ satisfies the equations
\begin{align}\label{QES}
    \mathcal{B}^{(t)}_{\mathcal{R}} +4 G \hbar\,\frac{\delta S}{\delta T} &= 0~,\\
    \mathcal{B}^{(x)}_{\mathcal{R}} +4 G \hbar\,\frac{\delta S}{\delta X} &= 0~,
\end{align}
where $(\mathcal{B}^{(t)}_{\mathcal{R}})$ and $(\mathcal{B}^{(x)}_{\mathcal{R}})$ denote the trace of the extrinsic curvature (expansion) in the two normal directions to $\mathcal{R}$, {\em i.e.}, $t^{\mu}$ and $x^{\mu}$ respectively. Similarly $\frac{\delta S}{\delta T}$ and $\frac{\delta S}{\delta X}$ are the entropy variations in the $t^{\mu}$ and $x^{\mu}$ directions respectively. 

The classical kink transform involves an extrinsic curvature shock at the classical RT surface. As shown in Sec.~\ref{kink}, extremality of the surface ensures that the constraint equations continue to be satisfied after the kink transform in this case. However, the quantum extremal surface has non-vanishing expansion, the constraint equations are not automatically satisfied when an extrinsic curvature shock is added at the quantum RT surface.

More precisely, the left hand side of the constraint equations on a slice $\Sigma$ are modified by the kink transform by
\begin{align}\label{constQ}
   \Delta \left(r_{\Sigma}-(K_{\Sigma})_{ab} (K_{\Sigma})^{ab} + (K_{\Sigma})^2 \right) &= 8 G \hbar \,\sinh(2\pi s) \,\frac{\delta S}{\delta T} \,\delta(X)~,\\
\Delta \left( D_{a} (K_{\Sigma})^{a}_{x} - D_{x} K_{\Sigma} \right)&= 4 G\hbar \,\sinh{(2\pi s)} \,\frac{\delta S}{\delta X} \,\delta(X)~,\\
\Delta \left( D_{a} (K_{\Sigma})^{a}_{i} - D_{i} K_{\Sigma} \right)&=0~,
\end{align}
where $\Delta$ represents the difference in the constraint equations between the original spacetime $\mathcal{M}$ and the kink transformed spacetime $\mathcal{M}_s$, and we have used Eqs.~\eqref{QES} and \eqref{onlychange}.
These are essentially the analogs of Eqs.~\eqref{econstraint} and \eqref{momentC}, and we have simplified the notation slightly.

For the constraint equations to be solved, the kink transform would have to generate the same change on the right hand side of the constraints. It would thus have to induce an additional stress tensor shock of the form
\begin{align}\label{shockQ}
    \Delta{T_{TT}}&=\frac{1}{2\pi}\sinh(2\pi s) \,\frac{\delta S}{\delta T}\,\delta(X)~,\\
    \Delta{T_{TX}}&=\frac{1}{2\pi}\sinh(2\pi s) \,\frac{\delta S}{\delta X}\,\delta(X)~.
\end{align}
Formally, these conditions agree precisely with the properties of CC flow discussed in Sec.~\ref{CC}. Thus, we might expect a generalized bulk CC flow to result in shocks of precisely this form.

In fact, the existence of semiclassical states satisfying the above equations was conjectured in \cite{Bousso:2019dxk}; the fact that CC flow generates such states in the non-gravitational limit was interpreted as non-trivial evidence in support of the conjecture. Thus, we expect a kink transform at the quantum extremal surface with a suitable modification of the state to provide the bulk dual of CC flow to all orders in $G$.

At a more speculative level, we can also discuss the bulk dual of CC flow in certain special states called fixed area states, which serve as a natural basis for modular flow \cite{Akers:2018fow,Dong:2018seb,Dong:2019piw}. These are approximate eigenstates of the area operator and are therefore unlike smooth semiclassical states which are analogous to coherent states. The Lorentzian bulk dual of such states potentially involves superpositions over geometries \cite{Marolf:2018ldl}.

However, by construction, the reduced density matrix is maximally mixed at leading order in $G$. Thus, the state $\ket{\psi}$ is unaffected by one sided modular flow, and the only effect of CC flow is that we describe the state on a kinematically related slice $\mathcal{C}_s$. Thus, the dual description must be invariant under CC flow up to a diffeomorphism. 
% Importantly, the constraint equations are relaxed at the RT surface for fixed area eigenstates, and hence, we do not face the issues addressed above.

In such states, one could apply the semiclassical prescription using Eq.~\eqref{JLMS}. As discussed above, the action of the area operators results in a diffeomorphism of the geometric description, if it exists. From Eq.~\ref{JLMS}, the remaining action of the boundary CC flow is to simply induce a bulk CC flow.

\subsection{Beyond Flat Cuts}

\begin{figure}[t]
\begin{center}
  \includegraphics[width=0.9\textwidth]{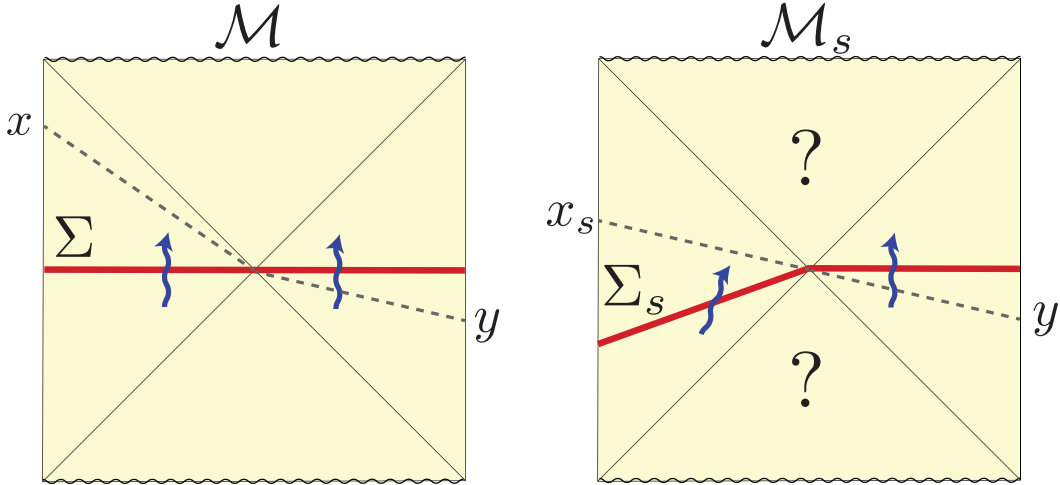} 
\end{center}
\caption{An arbitrary spacetime $\mathcal{M}$ with two asymptotic boundaries is transformed to a physically different spacetime $\mathcal{M}_s$ by performing a kink transform on the Cauchy slice $\Sigma$. A piecewise geodesic (dashed gray line) in $\mathcal{M}$ connecting $x$ and $y$ with boost angle $2\pi s$ at $\mathcal{R}$ becomes a geodesic between $x_s$ and $y$ in $\mathcal{M}_s$.}
\label{fig:twosided}
\end{figure}

Kink transform/CC flow duality can be generalized to other choices of boundary subsystems, so long as a precursor slice can be defined. The precursor slice is generated by acting on the original slice with the vacuum modular Hamiltonian; this is well-defined only if this action is geometric.
% Importantly, we also needed a Killing symmetry that preserves the entangling surface $\partial A$.
% This is because the kink transform automatically leads to an intrinsic geometry of $\mathcal{C}_s$ that is identical to $\mathcal{C}$.
In Sec.~\ref{precursor}, we ensured this by taking the boundary to be Minkowski space and choosing a planar cut. Precursor slices also exist in any conformally related choice, such as a spherical cut. 

But there are other settings where the vacuum modular Hamiltonian acts geometrically. This includes multiple asymptotically AdS boundaries where the boundary manifold has a time translation symmetry. For example, consider a two-sided black hole geometry $\mathcal{M}$ with a compact RT surface $\mathcal{R}$ as seen in Fig.~\ref{fig:twosided}.
The boundary manifold is of the form $\mathcal{C}\times\mathbb{R}$, where the first factor corresponds to the spatial geometry and the second corresponds to the time direction. The boundary Hilbert spaces factorizes; each boundary algebra is a Type-I factor.
Thus, the version of CC flow defined in terms of density matrices in Eq.~\eqref{eq:CCrho} becomes rigorous in this situation.
A natural choice of vacuum state is the thermofield double~\cite{Israel:1976ur,Maldacena:2001kr}. The reduced state on each side is thermal, $\rho_{A_0}\sim\exp(-\beta H)$. Thus the modular Hamiltonian is proportional to the ordinary Hamiltonian on each boundary. This generates time translations and so is geometric.

Now, in any such geometry $\mathcal{M}$ one can pick a Cauchy slice $\Sigma$ that ends on boundary time slices on both sides and contains $\mathcal{R}$. 
In obvious analogy with Sec.~\ref{kink}, we conjecture that the domain of dependence of the kink transformed slice $\Sigma_{s}$ in a modified geometry $\mathcal{M}_s$ is dual to the boundary state:
\begin{align}\label{twoside}
\ket{\psi_s(\mathcal{C}_s)}_{LR}= \rho_{L}^{-i s} \ket{\psi(\mathcal{C})}_{LR}~,
\end{align}
where we have used the notation of Eq.~\eqref{precursor-rho}.

In such a situation, it is again manifest that the Wheeler-DeWitt patches dual to either side are preserved by arguments similar to those made in Sec.~\ref{MLR}.
However, since there is no portion of $\mathcal{R}$ that reaches the asymptotic boundary, there is no analog of the shock matching done in Sec.~\ref{shocks}.
Notably, since $\partial A=\varnothing$ in this case, there is no subtlety regarding boundary conditions for JLMS and thus, one-sided modular flow makes sense without any regulator.
Thus, our construction is simply kinematically related to the construction in \cite{Faulkner:2018faa}.

%Other situations involving local modular Hamiltonians are subregions related to flat cuts by a conformal transformation.
%These are trivially related to our kink transform construction by a PBH diffeomorphism in the bulk that implements a conformal transformation on the boundary \cite{Imbimbo:1999bj}.

An interesting situation arises for wiggly cuts of the Rindler horizon, {\em i.e.}, $u=0$ and $v=V(y)$. The modular Hamiltonian acts locally, but only when restricted to the null plane \cite{Casini:2017roe}. Its action becomes non-local when extended to the rest of the domain of dependence. The properties of CC flow described in Sections \ref{general}-\ref{stsc} all hold for this choice of cut. This constrains one-sided operator expectation values on the null plane, subregion entropies for cuts entirely to one side of $V(y)$, and even the $T_{vv}$ shock at the cut. Interestingly, all of them are matched by the kink transform, by the arguments given in Sec.~\ref{kink}. Even the expected stress tensor shock can still be derived, by taking a null limit of our derivation as described in Appendix~\ref{nulllimit}. One might then guess that the kink transform is also dual to CC flow for arbitrary wiggly cuts.

Even in the vacuum, however, the kink transform across a wiggly cut results in a boundary slice that cannot be embedded in Minkowski space, due to the absence of a boost symmetry that preserves the entangling surface. Thus, the kink transform would have to be modified to work for wiggly cuts. The transformation of boundary observables off the null plane is quite complicated for wiggly cut CC flow. Thus, we also expect that regions of the entanglement wedge probed by such observables should be drastically modified, unlike the case where the entangling surface is a flat Rindler cut.

However, the wiggly-cut boundary transformation remains simple for observables restricted to the null plane. Thus one could try to formulate a version of the kink transform on Cauchy slices anchored to the null plane on the boundary and the RT surface in the bulk. Perhaps a non-trivial transformation of the entanglement wedge arises from the need to ensure that the kink transformed initial data be compatible with corner conditions at the junction where the slice meets the asymptotic boundary \cite{Horowitz:2019dym}. We leave this question to future work.

\subsection{Other Probes of CC Flow}
\label{other}

In Sec.~\ref{duality}, we provided evidence for kink transform/CC flow duality. The preservation of the left and right entanglement wedges under the kink transform ensures that all one-sided correlation functions transform as required. It would be interesting to consider two sided correlation functions.
However, these do not change universally and are difficult to compute in general. In the bulk, this is manifested by the fact that the future and past wedges do not change simply and need to be solved for.

However, because of the shared role of the kink transform, we can take advantage of the modular toolkit for one-sided modular flow~\cite{Faulkner:2018faa}. Let $\ket{\widetilde{\psi}_s} =\rho_A^{-i\,s} \ket{\psi}$ be a family of states generated by one-sided modular flow as discussed in Sec.~\ref{sec:JLMS}. Then certain two sided correlation functions $\bra{\widetilde{\psi}_s} O(x)\,O(y) \ket{\widetilde{\psi}_s}$ can be computed as follows.

Suppose $O(x)$ is an operator dual to a ``heavy'' bulk field with mass $m$ such that $1/\ell_{\text{AdS}} \ll m \ll 1/\ell_{\text{P}},1/\ell_{\text{s}}$.
Correlation functions for such an operator can then be computed using the geodesic approximation,
\begin{align}
\langle O(x) O(y) \rangle \approx \exp(-m L)~,
\end{align}
where $L$ is the length of the bulk geodesic connecting boundary points $x$ and $y$. Now consider boundary points $x$ and $y$ such that there is a piecewise bulk geodesic of length $L(x,y)$ joining them in the spacetime dual to the state $\ket{\psi}$.

This kinked geodesic is required to pass through the RT surface of subregion $A$ with a specific boost angle $2\pi s$ as seen in Fig.~\ref{fig:twosided}. (This is a fine-tuned condition on the set of points ${x,y}$.) Since single sided modular flow behaves locally as a boost at the RT surface, it straightens out the kinked geodesic such that it is now a true geodesic in the spacetime dual to the state $\ket{\widetilde{\psi}_s}$.
Thus, we have
\begin{align}
    \bra{\widetilde{\psi}_s} O(x)\,O(y) \ket{\widetilde{\psi}_s}\approx \exp(-m L(x,y))~.
\end{align}

As discussed in Sec.~\ref{precursor}, the CC flowed state can equivalently be thought of as the single sided modular flowed state $\ket{\widetilde{\psi}_s}$ on a kinematically transformed slice $\mathcal{C}_s$.
Thus, the above rules can still be used to compute two sided correlation functions in the CC flowed state $\ket{\psi_s}= u_s \ket{\psi}$ as
\begin{align}
    \bra{\psi_s}O(x_s) O(y)\ket{\psi_s} \approx \exp(-m L(x,y))~,
\end{align}
where $x_s$ is the point related to $x$ by the vacuum modular flow transformation.

We also note that the shock matching performed in Sec.~\ref{duality} was a near boundary calculation. However, a bulk shock exists everywhere on the RT surface.
One could solve for the position of the RT surface to further subleading orders and relate the bulk shock to the boundary stress tensor. This would yield a sequence of relations that the stress tensor must satisfy in order to be dual to the kink transform. In general these relations may be highly theory-dependent, but it would be interesting to see if some follow directly from CC flow or make interesting universal predictions for CC flow in holographic theories.

\subsection{Higher Curvature Corrections}
\label{hcc}
In Sec.~\ref{duality}, we argued that the bulk kink transform in a theory of Einstein gravity satisfies properties consistent with the boundary CC flow.
However, this result is robust to the addition of higher curvature corrections in the bulk theory.
The preservation of the two entanglement wedges, {\em i.e.}, Eq.~\eqref{CCkink22}, is a geometric fact that remains unchanged.\footnote{Here we assume that the initial value formulation of Einstein gravity can be perturbatively adjusted to include higher curvature corrections despite the fact that a non-perturbative classical analysis of higher curvature theories is often problematic due to the Ostrogradsky instability \cite{Woodard:2015zca}.}

Further, the matching of the stress tensor shock crucially depended on two ingredients.
Firstly, Eq.~\eqref{gexp}, the holographic dictionary between the boundary stress tensor and the bulk metric perturbation and secondly, Eq.~\eqref{holoS}, the relation between the boundary entropy variation and the shape of the RT surface.
Both of these relations are modified once higher curvature corrections are included \cite{Faulkner:2013ica,Leichenauer:2018obf}.
However, it follows generally from dimension counting arguments that 
\begin{align}
    g_{ij}^{(d)} &= \eta_1 \frac{16\pi G}{ d }\langle T_{ij}\rangle ~,\\
    X^A_{(d)} &= -\eta_2 \frac{4 G}{d} \left. \frac{\delta S}{\delta X^A}\right|_{\mathcal{R}}~,
\end{align}
where $\eta_1$ and $\eta_2$ are constants that depend on the higher curvature couplings.
Using the first law of entanglement, it can be shown that in fact $\eta_1=\eta_2$ \cite{Faulkner:2013ica,Leichenauer:2018obf}.
Hence, the boundary stress tensor shock obtained from the kink transform is robust to higher curvature corrections.

\subsection{Holographic proof of QNEC}
\begin{figure}[t]
\begin{center}
  \includegraphics[scale=0.25]{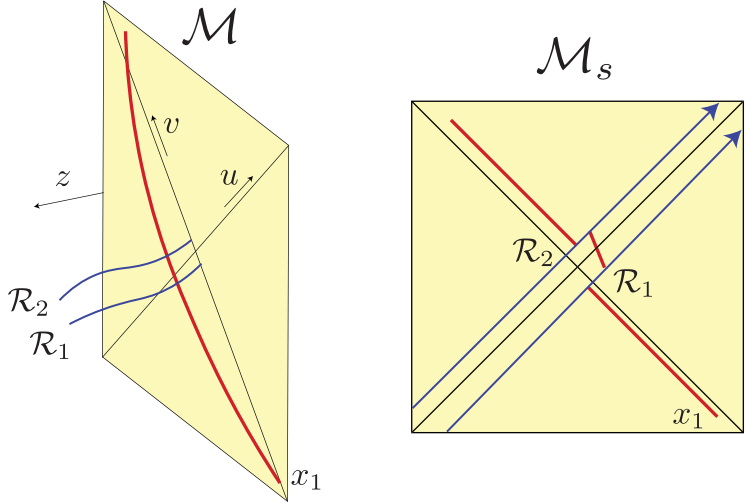} 
\end{center}
\caption{Holographic proofs. {\em Left:} Boundary causality is respected by the red curve that goes through the bulk in a spacetime $\mathcal{M}$; this is used in proving the ANEC. The RT surfaces $\mathcal{R}_1$ and $\mathcal{R}_2$ must be  spacelike separated; this is used in proving the QNEC. {\em Right:} In the kink transformed spacetime $\mathcal{M}_s$ as $s\rightarrow\infty$, the QNEC follows from causality of the red curve, which only gets contributions from the Weyl shocks (blue) at $\mathcal{R}_1$ and $\mathcal{R}_2$, and the metric perturbation in the region between them.
}
\label{fig:QNEC}
\end{figure}

A recent proof of the QNEC from the ANEC~\cite{Ceyhan:2018zfg} considers CC flow for a subregion $A$ on the null plane $u=0$ with entangling surfaces $v=V_1(y)$ and $v=V_2(y)$ surrounding a given point $p$. From the transformation properties of the stress tensor under CC flow described in Sec.~\ref{nullcuts}, $T_{vv}\rightarrow 0$ as $s\rightarrow\infty$. In addition, there are stress tensor shocks at $\partial A$, as described in Sec.~\ref{stsc}, of weight $f(s) \frac{\delta S}{\delta V(y)}\Big\lvert_{\psi,\partial A}$. In the limit $V_1(y)\rightarrow V_2(y)$, computing the ANEC in the CC flowed state, one obtains contributions from the stress tensor $T_{vv}(p)$ in subregion $A$, and a contribution proportional to $\frac{\delta^2 S}{\delta V(y_1)\delta V(y_2)}\Big\lvert_{\psi,p}$ from the shocks. Positivity of the averaged null energy in the CC flowed state then implies the QNEC in the original state.

Prior to the QFT proofs, both the ANEC and QNEC had been proved holographically \cite{Kelly:2014mra,Koeller:2015qmn}.
The guiding principle behind both of these proofs was the fact that consistency of the holographic duality requires bulk causality to respect boundary causality as we demonstrate in Fig.~\ref{fig:QNEC}.
In the case of the ANEC, one considers an infinitely long curve connecting points on past null infinity to future null infinity through the bulk and demands that it respect boundary causality \cite{Kelly:2014mra}.
In the proof of the QNEC, one requires that curves joining the RT surfaces of subregions $v<V_1(y)$ and $v>V_2(y)$, denoted $\mathcal{R}_1$ and $\mathcal{R}_2$, respect boundary causality \cite{Koeller:2015qmn, Balakrishnan:2017bjg}.
There are two contributions to the lightcone tilt of this bulk curve coming from the metric perturbation $h_{vv}$ in the near boundary geometry, and the shape of the RT surface $X^{\mu}(y,z)$.
By the holographic dictionary, these contributions can be related to the boundary stress tensor $T_{vv}$ and the entropy variations $\frac{\delta S}{\delta V}$ as discussed in Sec.~\ref{shocks}.

Now performing the kink transform removes the contribution coming from the shape of the RT surface and puts it into a time advance/delay coming from shocks in the bulk Weyl tensor that we compute in Appendix~\ref{nulllimit}.
Considering the extended curve from past to future null infinity, we see that whether or not it respects boundary causality is determined entirely by the region between the entangling surfaces $\mathcal{R}_1$ and $\mathcal{R}_2$ since the bulk solution approaches the vacuum everywhere else in the limit $s\rightarrow \infty$.
Requiring causality of the ANEC curve then results in the QNEC, making the connection to the boundary proof manifest.

\paragraph*{Acknowledgements}
We thank Chris Akers, Xi Dong, Thomas Faulkner, Daniel Jafferis, Ronak Soni, and Aron Wall for helpful discussions and comments.
This work was supported in part by the Berkeley Center for Theoretical Physics; by the Department of Energy, Office of Science, Office of High Energy Physics under QuantISED Award DE-SC0019380 and under contract DE-AC02-05CH11231; and by the National Science Foundation under grant PHY1820912.

\appendix

\section{Null Limit of the Kink Transform}\label{nulllimit}

In this appendix we apply the kink transform to a Cauchy slice $\Sigma$ that has null segments. In the null limit we express the kink transform in terms of the null initial value problem. We then show that this leads to a shock in the Weyl tensor for $d > 2$. From this Weyl shock we extract the boundary stress tensor shock. This serves a two-fold purpose. The first is that it provides direct intuition for how the kink transform modifies the geometry. The second is that, as will be evident from the calculation below, the derivation of the stress tensor shock from the Weyl shock works even for wiggly cuts of the Rindler horizon on the boundary.\footnote{The results of this section do not apply when $d=2$, as the shear and the Weyl tensor vanish identically. However in $d = 2$ there is no distinction between flat and wiggly cuts on the boundary so we gain neither additional intuition nor generality compared with the analysis in Sec.~\ref{shocks}.} 

Let $N_k$ be a null segment of $\Sigma$ in a neighborhood of $\mathcal{R}$ and let $k^{a}$ be the null generator of $N_k$. We now allow the boundary anchor of $\mathcal{R}$ to be an arbitrary cut $V_0(y)$ of the Rindler horizon, as considered in Sec.~\ref{nullcuts}. Lastly, denote by $P^a_{\mu}$ and $P^i_{\mu}$ the projectors onto $N_k$ and cross-sections of $N_k$ (including the RT surface $\mathcal{R}$), respectively. We can compose these to obtain the projector $P^i_a$.  

By Eq.~\eqref{onlychange}, when $\Sigma$ is spacelike in a neighborhood of $\mathcal{R}$ the kink transform can be contracted as follows:
\begin{align}\label{fornull}
x^{a} (K_{\Sigma})_{ab} \to x^{a} (K_{\Sigma})_{ab} - \sinh{(2\pi s)} x_{b} ~\delta(\mathcal{R})~.
\end{align}
In the null limit both $x^{a}$ and $t^{\mu}$ approach $k^{a}$. Therefore, the quantity in the LHS of Eq.~\eqref{fornull} has the following null limit:
\begin{align}
x^{a}(K_{\Sigma})_{ab} \stackrel{\text{null}}{\to} k^{a} \nabla_{a} k_{b}~.
\end{align}
The transformation of Eq.~\eqref{fornull} then becomes
\begin{align}\label{leftstretch}
\kappa\to \kappa - \sinh{(2\pi s)} \delta(\lambda)~,
\end{align}
where $\lambda$ is a null parameter adapted to $k^a$ and $\kappa$ is the inaffinity defined by 
\begin{align}
    k^b \nabla_b k^a = \kappa k^a~.
\end{align}
We refer to this transformation as the \emph{left stretch}, as it arises from a one-sided dilatation along $N_k$. This transformation was originally described in \cite{Bousso:2019dxk} in the context of black hole coarse-graining. 

We now show that the left stretch generates a Weyl tensor shock at the RT surface. The shear of a null congruence is defined by 
\begin{align}
    \sigma_{ij} = P^a_i P^b_j \nabla_{(a}k_{b)}~.
\end{align}
It satisfies the evolution equation \cite{Gourgoulhon:2007ue}
\begin{align}
    \mathcal{L}_k \sigma_{ij} = \kappa \sigma_{ij} + \sigma_{i}{}{}^k \sigma_{kj} - P_i ^{\mu} P_j^{\mu} k^a k^b C_{ a \mu b \nu} ~.
\end{align}
Now let $\lambda$ be a parametrization of $N_k$ adapted to $k^a$, with $\lambda =0$ corresponding to $\mathcal{R}$. In terms of $\lambda$, the evolution equation can be written as 
\begin{align}
    \partial_{\lambda}\sigma_{ij} = \kappa \sigma_{ij} + \sigma_i{}{}^k \sigma_{kj} - C_{\lambda i \lambda j} ~. \label{beforeshock}
\end{align}

Consider now the new spacetime $\mathcal{M}_s$ generated by the left stretch. As in Sec.~\ref{shocks}, we denote quantities in $\mathcal{M}_s$ with tildes. We can then write the evolution equation in $\mathcal{M}_s$,  
\begin{align}
     \partial_{\tilde \lambda}\tilde \sigma_{ij} = \tilde\kappa \tilde\sigma_{ij} + \tilde\sigma_i{}{}^k\tilde \sigma_{kj} - \tilde C_{\lambda i \lambda j} ~. \label{aftershock}
\end{align}
Since $k^a$ is tangent to $N_k$, and $(N_k)_s = N_k$ as submanifolds, we can identify $k^a$ with $\tilde k^a$. Thus we can use the same parameter $\lambda$ in both spacetimes. Since $\sigma_{ij}$ is intrinsic to $N_k$, we can identify $\sigma_{ij}$ and $\tilde \sigma_{ij}$ for the same reason. Comparing Eqs.~\eqref{beforeshock} and \eqref{aftershock}, and inserting Eq.~\eqref{leftstretch}, we find that there is a Weyl shock 
\begin{align}
    \tilde C_{\lambda i \lambda j} = C_{\lambda i \lambda j}-\sinh(2\pi s)\sigma_{ij}\delta(\lambda) ~. \label{Weyl}
\end{align}

We now show that the Weyl shock Eq.~\eqref{Weyl} reproduces the near boundary shock Eq.~\eqref{nullshock}, but now for wiggly cuts of the Rindler horizon. To do this, we evaluate both $\sigma_{ij}$ and $C_{\lambda i \lambda j}$ in Fefferman-Graham coordinates to leading non-trivial order. The Fefferman-Graham coordinates for $\mathcal{M}$ and $\mathcal{M}_s$ are defined exactly as in Sec.~\ref{shocks}, except we now use null coordinates $(u,v)$ and $(\tilde u, \tilde v)$ on the boundary as defined in Sec.~\ref{precursor}. To start with, we note that $k_a\partial_z \bar{X}^a =0$ since $\partial_z \bar{X}^a$ is tangent to the RT surface. Evaluating this at leading order yields the relation 
\begin{align}
    k_z = -dz^{d-3}\mathcal{U}_{(d)} + \mathcal{O}(z^{d-4})~. \label{zcomp}
\end{align}
We recall that
\begin{align}
    \mathcal{U}_{(d)} = -\frac{4G}{d}\frac{\delta S}{\delta V}\Big \lvert_{V_0}~.
\end{align}
Moreover, the projector is given by  
\begin{align}
    P^{\mu}_i = \partial_i \bar{X}^{\mu} ~.
\end{align}
From this definition, one can check that 
\begin{align}
   P_i^{z} &= \delta_i^z + \mathcal{O}(z^{d-1}) ~,\\
    P_i^A &= \mathcal{O}(z^{d-1}) ~.
\end{align}
Furthermore, 
\begin{align}
    \nabla_z k_A, &\ \nabla_A k_z \sim \mathcal{O}(z^{-1}) ~,\nonumber\\ 
    \nabla_A k_B &\sim \mathcal{O}(1) \nonumber ~, \\ 
    \nabla_z k_z &= -d(d-2)\mathcal{U}_{(d)}z^{d-4} + \mathcal{O}(z^{d-5})~,
\end{align}
where we have used that $k^A\sim \mathcal{O}(1)$. Hence to leading order we simply have 
\begin{align}
\sigma_{ij} = -d(d-2)\mathcal{U}_{(d)}z^{d-4}\delta_i^z \delta_j^z+\mathcal{O}(z^{d-5}) ~.
\end{align}
Finally, a straightforward but tedious calculation of the Weyl tensor yields 
\begin{align}
    \tilde C_{\tilde vi \tilde v j} =  C_{ v i v j} -8\pi G(d-2)\left(\langle \tilde{T}_{\tilde v \tilde v}\rangle - \langle T_{vv}\rangle\right) z^{d-4}\delta_i^z\delta_j^z \delta(\tilde v - V_0) + \mathcal{O}(z^{d-5})~,
\end{align}
where we have used that $\lambda \rightarrow v,\tilde{v}$ as $z\rightarrow 0$ in $\mathcal{M},\mathcal{M}_s$ respectively. Putting this together yields the desired shock for wiggly cuts of the Rindler horizon.

\bibliographystyle{JHEP}
\bibliography{leftstretch}

\end{document}